\documentclass[showpacs,aps,rmp,twocolumn,eqsecnum,longbibliography]{revtex4-1}
\usepackage{amssymb,amsmath,bm,graphicx}
\usepackage{multirow} 
\usepackage{footmisc} 

\usepackage{times}
\usepackage{CJKutf8}	

\usepackage{hyperref}
\hypersetup{%
pdftitle={Colloquium: Graphene spectroscopy},%
pdfauthor={D. N. Basov et al.},%
pdfpagemode={UseNone},%
pdfstartview={FitH},%
breaklinks=true,%
citecolor=blue,%
colorlinks=true,%
linkcolor=blue,%
urlcolor=blue}

\begin{document}
\begin{CJK*}{UTF8}{gbsn}	

\newcommand\UCSD{Department of Physics, University of California San Diego, 9500 Gilman Drive, La Jolla, California 92093, USA}
\newcommand\UCBA{Department of Physics, University of California at Berkeley, Berkeley, California 94720, USA}
\newcommand\UCBB{Materials Science Division, Lawrence Berkeley National Laboratory, Berkeley, California 94720, USA}
\newcommand\FU{State Key Laboratory of Surface Physics and Department of Physics, Fudan University, Shanghai 200433, China}

\title{\textit{Colloquium}: Graphene spectroscopy}

\author{D. N. Basov}
\author{M. M. Fogler}
\affiliation{\UCSD}

\author{A. Lanzara}

\author{Feng Wang}
\affiliation{\UCBA}
\affiliation{\UCBB}

\author{Yuanbo Zhang (\CJKchar{"5F}{"20}\CJKchar{"8F}{"DC}\CJKchar{"6C}{"E2})}
\affiliation{\FU}

\date{\today}

\begin{abstract}

Spectroscopic studies of electronic phenomena in graphene are reviewed.
A variety of methods and techniques are surveyed, from
quasiparticle spectroscopies (tunneling, photoemission) to
methods probing density and current response (infrared optics, Raman)
to scanning probe nanoscopy and ultrafast pump-probe experiments.
Vast complimentary information derived from these investigations is shown to highlight
unusual properties of Dirac quasiparticles
and many-body interaction effects in the physics of graphene.

\end{abstract}


\maketitle 
\end{CJK*} 

\tableofcontents

\section{Introduction} 
\label{sec:Introduction}

\subsection{Scope of this review} 

\begin{figure*}
\includegraphics[width=6.00in]{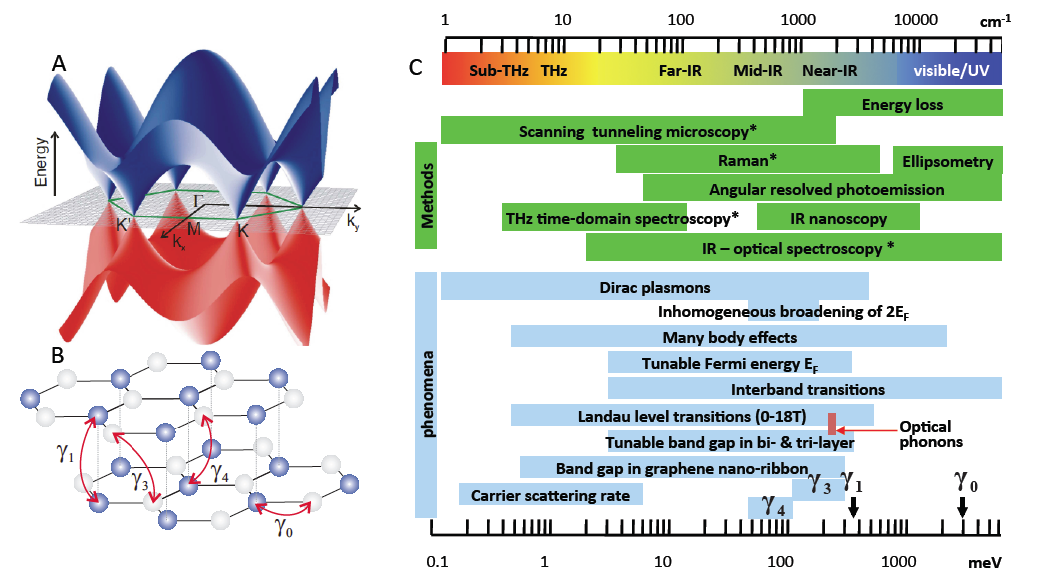}
\caption{\label{fig:1.2.1}
(Color online) Panel A: a schematic of the $\pi$-band dispersion of SLG showing Dirac cones at $\mathrm{K}$ and $\mathrm{K}'$ points.
After \cite{Orlita2010des}.
Panel B: the definitions of the intra ($\gamma_0$) and interlayer ($\gamma_1$--$\gamma_4$) hopping parameters of Bernal-stacked graphene materials.
[For their experimental values, see, e.g., \cite{Zhang2008dot}.] 
Panel C: the energy scales of electronic phenomena in graphene along with the corresponding frequency ranges and spectroscopic methods.
The asterisk (*) denotes compatibility of a method with high magnetic fields.
}
\end{figure*}

Graphene is a single atomic layer of $sp^2$-hybridized carbon atoms arranged in a honeycomb lattice. 
This two-dimensional (2D) allotrope of carbon is characterized by a number of superlative virtues \cite{Geim2009gsa}, e.g.,
a record-high electronic mobility at ambient conditions \cite{Morozov2008gic},
exceptional mechanical strength \cite{Lee2008mot}, and
thermal conductivity \cite{Balandin2008stc, Ghosh2008eht}
Remarkable properties of graphene have ignited tremendous interest
that resulted in approximately 50,000 publications at the time of writing.
A number of authoritative reviews\footnote{%
See \textcite{CastroNeto2009tep, Peres2010ctt, DasSarma2011eti, Kotov2012eei, Katsnelson2012gci, McCann2013epb}.
}
have been written to survey this body of literature
but no single review can any longer cover the entire topic.
The purpose of this Colloquium is to overview specifically the spectroscopic experiments
that have helped to shape the modern understanding of the physical properties of graphene.
While selected topics in graphene spectroscopy have been discussed,\footnote{%
See \textcite{Orlita2010des} for optics,
\textcite{Ni2008rsa, Dresselhaus2012rsc} for Raman scattering,
\hbox{\textcite{Li2012stm}} for scanning tunneling spectroscopy,
and \textcite{Connolly2010nog} for other scanned probes.}
here we aim to present a panoramic view of physical phenomena in graphene emerging from both spectroscopy and imaging (Fig.~\ref{fig:1.2.1}C).

Spectroscopic observables can be formally categorized as either
quasiparticle or current/density response functions.
The former are fermionic, the latter are bosonic.
The former is traditionally measured by photoemission and tunneling spectroscopy,
while the latter can be investigated by, e.g., optical spectroscopy.
Yet it may be possible to infer
both quasiparticle and collective properties from the same type of measurements.
For example,
fine anomalies of the quasiparticle spectra seen in photoemission
can give information about interactions
between quasiparticles and collective modes (Sec.~\ref{sec:e-ph-pl})
Conversely, optical conductivity, which is a collective response,
enables one to infer, with some approximation,
the parameters of a quasiparticle band-structure (Secs.~\ref{sec:direct}, \ref{sec:renormalization}, \ref{sec:Landau}, and \ref{sec:BLG}).

Finding such connections is facilitated by
spectacular tunability of graphene.
For example, with photoemission or tunneling techniques one can
monitor the chemical potential $\mu$ of graphene as a function of the electron concentration $N$ and
thereby extract the thermodynamic density of states.
The same physical quantity can be measured by a very different technique, the scanning
single-electron transistor microscopy.
In our analysis of such complementary data we focus on what we believe are the most pressing topics in the physics of graphene, e.g., many-body effects.
Additionally, our review covers 
information obtained by scanned probes and out-of-equilibrium methods
that greatly expand available means to study graphene in space and time domains.
Finally, we briefly address phenomena that arise when physical properties of graphene
are altered via its environment and nanostructuring.


\subsection{Graphene morphology} 
\label{sec:forms}

Graphene can be isolated or fabricated in a number of different forms, which is an important consideration in spectroscopy.
Effectiveness of a given spectroscopic tool depends on
the accessibility of the sample surface to the incident radiation.
The size of the accessible area must normally be larger than the wavelength of the incident beam
unless near-field probes are employed (Sec.~\ref{sec:plasmons})
Mosaic structure and defects may affect momentum and energy resolution of the measurement. 
Graphene differs widely in terms of these parameters depending on preparation method.
Mechanical exfoliation of graphite typically produces single, bi-, and multi-layer graphene (SLG, BLG, and MLG, respectively) of a few $\mu\mathrm{m}$ in size,
although occasionally samples of dimensions of hundreds of $\mu\mathrm{m}$ can be obtained.
Exfoliated samples can be transferred onto insulating substrates,
after which they can be gated and subject to transport measurements.
The sign and the magnitude of carrier concentration $N$ in gated samples can be precisely controlled over a wide range.
The lower bound on $|N| \sim 10^{10}\,\mathrm{cm}^{-2}$ is set by inhomogeneities (Sec.~\ref{sec:Inhomogeneities}).
The upper bound $|N| \sim 10^{13}\,\mathrm{cm}^{-2}$
is limited by the dielectric breakdown strength of the substrate,
although still higher $|N|$ are achievable by electrolytic gating .\footnote{See \textcite{Mak2009ooa, Xia2009moq, Efetov2010cep, Ju2011gpf, Newaz2012pcs}.}
The carrier concentration can also be controlled by doping \cite{Chen2008cis}.

Morphologically, exfoliated samples are single crystals.
They hold the record for transport mobility $\mu_{tr}$ although it varies much with the type of the substrate.
Currently, high-quality hexagonal boron nitride (hBN) substrates enable one to achieve $\mu_{tr} \sim 10^5\,\mathrm{cm}^2/\mathrm{Vs}$,
which is about an order of magnitude higher than what is typical for graphene on SiO$_2$
and corresponds to $\mu\mathrm{m}$-scale mean-free path \cite{Dean2010bns, Mayorov2011msb}.
The highest mobility $\sim 10^6\,\mathrm{cm}^2/\mathrm{Vs}$ is demonstrated by exfoliated graphene that is suspended off a substrate and subject to current annealing \cite{Du2008abt, Bolotin2008uem, Elias2011dcr}.
Mechanical instabilities limit the size of suspended devices to $1$--$2\,\mu\mathrm{m}$ and restrict the maximum $|N|$ to a few times $10^{11}\,\mathrm{cm}^{-2}$.

Large-area graphene can be made by another method: epitaxial growth on SiC by thermal desorption of Si \cite{vanBommel1975laa}.
Epitaxial graphene may contain a single or many dozens of layers.
The initial layer (layer number $L = 0$) has strong covalent bonds to the SiC substrate and is electronically different from the ideal SLG \cite{deHeer2007eg}.
The morphology and electron properties of the subsequent layers, $L > 0$,
depend on which SiC crystal face it is grown: the Si-terminated $(0001)$ face or the C-terminated $(000\bar{1})$ face.\footnote{See \textcite{Berger2004ueg, Berger2006eca, Forbeaux1998hgo, Charrier2002ssd, Nagashima1993eso, Rollings2006sac, Ohta2006ces, Emtsev2009tws}.}
According to \textcite{deHeer2011laa},
the Si-face grown graphene is orientationally ordered and has the Bernal stacking (as in graphite). The structure of the C-face epitaxial graphene is consistent with a stacking where every other layer is rotated by approximately $\pm 7^\circ$ with respect to a certain average orientation.
The rotations inhibit interlayer tunneling so that the band structure of each layer is similar to SLG (see also Sec.~\ref{sec:substrate}).

The morphology of the epitaxial graphene after annealing resembles a carpet draping over the staircase \cite{Emtsev2009tws}.
It is characterized by domains
a few $\mu\mathrm{m}$ wide and up to $50\,\mu\mathrm{m}$ long that mirror the underlying SiC terraces \cite{Emtsev2009tws, deHeer2011laa}.
%

The graphene/SiC interface is charged, inducing the $n$-type doping of about $10^{13}\,\mathrm{cm}^{-2}$
in the first ($L = 1$) graphene layer. 
Other layers have much smaller carrier concentration because of screening.
The screening length of about one layer was measured by ultrafast infrared (IR) spectroscopy \cite{Sun2010smo}.
The doping of the surface layers can be altered by depositing charged impurities \cite{Ohta2006ces, Zhou2008mti}.
Relatively low mobility $\mu_{tr} = 500$--$10,000\,\mathrm{cm}^2/\mathrm{Vs}$, the inhomogeneity of the doping profile, and the lack of its \textit{in situ} control can be seen as drawbacks of (the first generation of) epitaxial compared to exfoliated graphene. On the other hand, the much larger surface area of the epitaxial graphene is advantageous for spectroscopic studies and applications \cite{deHeer2007eg}.
An important recent breakthrough is
epitaxial growth of graphene on high-quality hBN substrates 
\cite{Yang2013egs}.

Graphene samples of strikingly large 30-in width \cite{Bae2010rtr} can be produced by the chemical vapor deposition (CVD) on metallic surfaces, e.g., Ru,%
Ni
or Cu
that act as catalysts.
CVD graphene can be transferred to insulating substrates making it amenable to gating and transport experiments \cite{Kim2009lsp, Bae2010rtr}.
The microstructure of CVD graphene sensitively depends on the roughness of the metallic substrate and the growth conditions.
Typical structural defects of CVD graphene are wrinkles and folds
induced by transfer process and also by thermal expansion of graphene upon cooling.
%
%
Grain boundaries are other common defects that have been directly imaged by micro-Raman \cite{Li2010gfw},
transmission electron microscopy \cite{Huang2011gag},
scanning tunneling microscopy \cite{Tapaszto2012mep, Koepke2013ase},
and near-field microscopy \cite{Fei2013epp}.
The corresponding domain sizes range between $1$--$20\,\mu\mathrm{m}$.
On the other hand,
graphene single crystals with dimension $\sim 0.5\,\mathrm{mm}$ have been grown on Cu by CVD \cite{Li2011lag}.
Transport mobilities of CVD-grown graphene and epitaxial graphene on SiC are roughly on par.

At the opposite extreme of spatial scales are nanocrystals and nanoribbons.
Graphene crystals of nm-size can be synthesized by reduction of graphene oxide\footnote{This can be done chemically \cite{Boehm1962, Dikin2007pac} or via IR irradiation \cite{El-Kady2012lso}.}
or by
ultrasonic cleavage of graphite in an organic solvent \cite{Hernandez2008hyp, Nair2012shp}.
Laminates of such crystals can be of macroscopic size amenable to X-ray and Raman spectroscopy.
Nanocrystals can also be grown epitaxially on patterned SiC surface \cite{deHeer2011laa}.
Graphene nanoribbons (GNRs) can be produced by
lithography, nanoparticle etching, and unzipping of carbon nanotubes.
%
%
%
There have been a number of spectroscopic studies of GNRs by scanned probes  \cite{Tao2011sre}, transport\footnote{See
\textcite{Han2007ebg, Liu2008eco, Todd2008qdb, Stampfer2009egi, Oostinga2010mtg, Gallagher2010dig, Han2010eti}.},
and photoemission \cite{Siegel2008sde, Zhou2008oot}
but because of space limitations they could not be covered in this review.

\subsection{Electronic structure of graphene neglecting interactions} 
\label{sec:single}

In this section we summarize basic facts about the SLG band-structure within the independent electron approximation \cite{CastroNeto2009tep}.
The nearest-neighbor carbon atoms in SLG form $sp^2$ bonds, which
give rise to the $\pi$ and $\sigma$ electron bands.
The $\sigma$-bands are relevant mostly for electronic phenomena at energies $\gtrsim 3\,\text{eV}$.
The unique low-energy properties of graphene derive from the $\pi$-bands
whose structure can be understood within the tight-binding model \cite{Wallace1947tbt}.
If only the nearest-neighbor transfer integral $\gamma_0=3.0\pm 0.3\,\text{eV}$ (Fig.~\ref{fig:1.2.1}B) is included,
the amplitudes $\psi_j$ of the Bloch functions on the two triangular sublattices $j = \mathrm{A}$ or $\mathrm{B}$ of the full honeycomb lattice
can be found by diagonalizing the $2 \times 2$ Hamiltonian
\begin{equation}
H_{\text{SLG}} =
\begin{pmatrix}
E_D                                                 & -{\gamma }_0 S_{\mathbf{k} } \\ 
-{\gamma }_0 S^*_{\mathbf{k} } & E_D
\end{pmatrix}\,,
\label{eqn:H_SLG}
\end{equation}
where $E_D$ is the constant on-site energy,
$\mathbf{k}  = \left(k_x, k_y\right)$ is the in-plane crystal momentum,
$S_{\mathbf{k} } = \exp({i k_x a / \sqrt{3}}) + 2 \exp({-i k_x a / 2\sqrt{3}})
\cos (k_y a / 2)$ represents the sum of the hopping amplitudes between a given site and its nearest neighbors, and $a = 2.461\,$\AA\ is the lattice constant.
The spectrum of $H_{\text{SLG}}$ has the form $\varepsilon_\pm(\mathbf{k} ) = E_D \pm \gamma_0 \left|S_{\mathbf{k} }\right|$ or
\begin{equation}
\label{eqn:E} 
\varepsilon_\pm = E_D \pm \gamma_0 \sqrt{3 + 2 \cos k_y a
 + 4 \cos\frac{\sqrt{3} k_x a}{2}\, \cos\frac{k_y a}{2}}\,.
\end{equation}
At energies $|\varepsilon - E_D| \ll \gamma_0$, this dispersion has an approximately conical shape $\varepsilon_\pm(\mathbf{k} ) = E_D \pm \hbar v_0 |\mathbf{k}  - \mathbf{K}|$
with velocity
$
v_0 = \frac{\sqrt{3}}{2}\, \frac{\gamma _0}{\hbar}\, a
= 0.9\text{--}1.0 \times 10^8\,\text{cm} / \text{s}
$
near the corners of the hexagonal Brillouin zone (BZ), see Fig.~\ref{fig:1.2.1}A\@.
Only two of such corners are inequivalent, e.g., $\mathbf{K},\: \mathbf{K}' = \bigl(\frac{2\pi}{\sqrt{3}\, a}, \pm \frac{2\pi}{3 a}\bigr)$; the other four are obtained via reciprocal lattice translations.
Near the $\mathbf{K}$ point, $H_{\text{SLG}}$ can be expanded to the first order in $q_\parallel$ and $q_\perp$ --- the components of vector $\mathbf{q} = \mathbf{k}  - \mathbf{K}$ parallel and perpendicular to $\mathbf{K}$, respectively.
This expansion yields the 2D Dirac Hamiltonian
\begin{equation}
H = E_D + \hbar v_0 (q_\parallel \sigma_x + q_\perp \sigma_y)\,,
\label{eqn:Dirac}
\end{equation}
which prompts analogies between graphene and quantum electrodynamics \cite{Katsnelson2007gnb}.
Here $\sigma_x$, $\sigma_y$ are the Pauli matrices.
Expansion near $\mathbf{K}'$ points gives a similar expression except for the sign of the $q_\parallel$-term.
The eigenvector $\Psi = (\psi_A, \psi_B)^{\mathrm{T}}$ of $H$ can be thought of as a spinor.
The direction of the corresponding pseudospin is parallel (antiparallel) for energy $\varepsilon_+$
($\varepsilon_-$).
The definite relation between the pseudospin and momentum directions is referred to as the \emph{chirality}.

The conical dispersion
yields the single-particle density of states (DOS) $\nu(E)$ linear in $|E - E_D|$.
Accounting for the four-fold degeneracy due to spin and valley,
one finds
\begin{equation}
\nu(E) = \frac{2}{\pi \hbar^2 v_0^2}\, |E - E_D|\,.
\label{eqn:nu_0}
\end{equation}
The frequently needed relations between the zero-temperature chemical potential $\mu$ (referenced to the Dirac point energy $E_D$), Fermi momentum $k_F$, and the carrier density $N$ read:
\begin{equation}
k_F = \sqrt{\pi |N|\,}\,,
\quad
\mu \equiv E_F - E_D = \mathrm{sign}(N)\, \hbar v_0 k_F\,.
\label{eqn:k_F}
\end{equation}
For $E - E_D$ not small compared to $\gamma_0$, deviations from the simplified Dirac model arise.
The spectrum exhibits saddle points at energies $E_D \pm \gamma_0$, which are reached at the three inequivalent points of the BZ: $\mathbf{M} = \bigl(\frac{2\pi}{\sqrt{3}\, a}, 0\bigr)$ and $\mathbf{M}', \mathbf{M}'' = \bigl(-\frac{\pi}{\sqrt{3}\, a}, \pm\frac{\pi}{a}\bigr)$, see Fig.~\ref{fig:1.2.1}A\@.
%
%
The DOS has logarithmic van~Hove singularities at these saddle-points.
In the noninteracting electron picture,
direct ($q = 0$) transitions between the conduction and valence band states of a given saddle-point would yield resonances at the energy $\hbar\omega = 2\gamma_0 \approx 5.4\,\mathrm{eV}$.
(Actually observed resonances are red-shifted due to interaction effects, see Sec.~\ref{sec:Excitations}.)

\subsection{Many-body effects and observables}
\label{sec:interaction}

While the single-electron picture is the basis for our understanding of electron properties of graphene,
it is certainly incomplete.
One of the goals of the present review is to summarize spectroscopic evidence for
many-body effects in graphene.
In this section we introduce the relevant theoretical concepts.
For simplicity, we assume that the temperature is zero and neglect disorder.

The strength of Coulomb interaction $U(r) = e^2 / \kappa r$ in graphene is controlled by the ratio
\begin{equation}
\alpha = \frac{e^2}{\kappa \hbar v_0}\,,
\label{eqn:alpha}
\end{equation}
where $\kappa$ is the effective dielectric constant of the environment.
Assuming $v_0 \approx 1.0 \times 10^8\,\mathrm{cm}/\mathrm{s}$,
for suspended graphene ($\kappa = 1$) one finds $\alpha \approx 2.3$, so that the interaction is quite strong.
Somewhat weaker interaction $\alpha \approx 0.9$ is realized for graphene on the common SiO$_2$ substrate, $\kappa = (1 + \epsilon_{\mathrm{SiO}_2}) / 2 = 2.45$.
For graphene grown on metals the long-range part of the interaction is absent,
with only residual short-range interaction remaining.

In general, spectroscopic techniques measure either quasiparticle or current (density) response functions.
Within the framework of the Fermi-liquid theory \cite{Nozieres1999toq}
interactions renormalize the quasiparticle properties,
meaning they change them quantitatively.
The current/density response is altered qualitatively due to emergence of collective modes.

A striking theoretical prediction made two decades ago, \textcite{Gonzalez1994nfl}
is that Coulomb interaction among electrons should cause a logarithmically divergent
renormalization of the Fermi velocity in undoped SLG,
\begin{equation}
\frac{v(q)}{v(k_c)} = 1 + \frac14 \alpha(k_c) \ln \frac{k_c}{q}
\:\:\text{at}\:\: k_F = 0\,,
\label{eqn:v_k_RG}
\end{equation}
which implies the negative curvature of the ``reshaped'' Dirac cones \cite{Elias2011dcr}.
Here, $k_c$ is the high momentum cutoff and $q = |\mathbf{k} - \mathbf{K}|$ is
again the momentum counted from the nearest Dirac point $\mathbf{K}$.
The physical reason for the divergence of $v(q)$
is the lack of metallic screening in undoped SLG because of vanishing thermodynamic density of states (TDOS) $\nu_T = {d N} / {d (\mu + E_D)}$.

While Eq.~\eqref{eqn:v_k_RG} can be obtained from the first-order
perturbation theory \cite{Barlas2007cac, Hwang2007dde, Polini2007gap},
the renormalization group (RG) approach of \textcite{Gonzalez1994nfl} indicates
that validity of this equation extends beyond the
weak-coupling case $\alpha \ll 1$.
It remains valid even at $\alpha \sim 1$
albeit in the asymptotic low-$q$ limit where the \textit{running} coupling constant
$\alpha(q) \equiv {e^2} / {\kappa \hbar v(q)} \ll 1$
is small.
The RG flow equation underlying Eq.~\eqref{eqn:v_k_RG},
\begin{equation}
\beta(\alpha) \equiv \frac{d \ln\alpha}{d \ln q} \simeq \frac{\alpha}{4}\,,
\quad \alpha \ll 1\,,
\label{eqn:beta_1}
\end{equation}
is free of nonuniversal quantities $\kappa$ and $k_c$,
and so in principle it can be used to compare the renormalization effects in different graphene materials.
The problem is that the asymptotic low-$q$ regime is hardly accessible in current experiments where
one typically deals with the nonperturbative case $\alpha \sim 1$.
Theoretical estimates \cite{Gonzalez1999mfl,Son2007qcp, Foster2008gvl} of the $\beta$-function in this latter regime yield
\begin{equation}
\beta \approx 0.2\,,
\quad \alpha \sim 1.
\label{eqn:beta_N}
\end{equation}
The corresponding renormalized velocity scales as
\begin{equation}
v(q) \sim q^{-\beta}\,.
\label{eqn:v_power-law}
\end{equation}
Distinguishing this weak power law from the logarithmic one~\eqref{eqn:v_k_RG} would still require a wide range of $q$.

The gapless Dirac spectrum should become unstable once $\alpha$ exceeds some critical value \cite{Drut2009igi, Khveshchenko2001gei, Sheehy2007qcs}.
It is unclear whether this transition may occur in SLG
as no experimental evidence for it has been reported.

In doped SLG the RG flow~\eqref{eqn:beta_1} is terminated at the Fermi momentum scale.
Therefore, velocity renormalization should be described by
the same formulas as in undoped one at $q \gg k_F$ but
may have extra features at $q \leq k_F$.
This expectation is born out by calculations \cite{DasSarma2007mbi}.
The result for the Fermi velocity, written in our notations, is
\begin{equation}
\frac{v_F}{v(k_c)} = 1 + \frac{\alpha}{\pi } \left(\ln \frac{1}{\alpha} - \frac{5}{3}\right)
+ \frac{\alpha}{4} \ln \frac{k_{c}}{k_{F} }\,,
\quad \alpha \ll 1\,,
\label{eqn:v_F_doped}
\end{equation}
where $\alpha$ should be understood as $\alpha(k_F)$.
Comparing with Eq.~\eqref{eqn:v_k_RG},
we see that $v_F$ is larger than $v(q)$ in undoped SLG at the same momentum $q = k_F$ by
an extra logarithmic term $\sim \alpha |\ln \alpha|$.
This logarithmic enhancement of the Fermi velocity is generic for an electron gas
with long-range Coulomb interactions in any dimension \cite{Giuliani2005qto}.
As a result, the renormalized dispersion has an inflection point near $k_F$ [see, e.g., \cite{Principi2012ttd, DasSarma2013vra}] and a positive (negative) curvature at smaller (larger) $q$.

Renormalization makes the relation between observables and quasiparticle properties such as $v(q)$ more complicated than in the noninteracting case.
For illustration, consider three key spectroscopic observables:
the single-particle DOS $\nu(E)$, the TDOS $\nu_T$,
and the threshold energy $\hbar \omega_{th}$ of interband optical absorption.
Since for curved spectrum phase and group velocities are not equal,
we must first clarify that by $v(q)$ we mean the latter, i.e.,
the slope of the dispersion curve $E(q)$.
In theoretical literature, $E(q)$ is usually defined by the equation
\begin{equation}
E(q) = \varepsilon(q) + \Sigma_1\bigl(q, E(q)\bigr)\,,
\label{eqn:E_Dirac}
\end{equation}
where  $\Sigma(q,\omega) = \Sigma_1(q , \omega) + i \Sigma_2(q , \omega)$
is the electron self-energy and the subscripts $\pm$ are suppressed to lighten the notations.
In experimental practice (Sec.~\ref{sec:Dirac}),
more directly accessible than $\Sigma(q,\omega)$ is the spectral function
\begin{equation}
A(q, \omega)
= \frac{-2 \Sigma_2(q, \omega)}
      {[\omega - \varepsilon(q) - \Sigma_1(q, \omega)]^2 + [\Sigma_2(q, \omega)]^2}\,,
\label{eqn:A}
\end{equation}
and the more convenient definition of $E(q)$ is the energy $\omega$ at which $A(q, \omega)$ has a maximum.
As long as this maximum is sharp so that the quasiparticles are well-defined,
the two definitions are equivalent.
For the velocity, they entail
\begin{equation}
\frac{v(q)}{v_0} \equiv \frac{1}{\hbar v_0} \frac{d E}{d q}
= \left(1 + \frac{\partial_q \Sigma_1}{\hbar v_0}\right) Z(q)\,.
\label{eqn:v_q}
\end{equation}
%
The three quantities in question, $\nu$, $\nu_T$, and $\hbar\omega_{th}$, are related to $v(q)$ as follows:
\begin{align}
\nu(E) &\simeq \frac{2}{\pi} \frac{q}{\hbar v(q)} Z(q)\,,
\quad
Z(q) \equiv \frac{1}{1 - \partial_E \Sigma_1}\,,
\label{eqn:nu_E}\\
\nu_T(N) &\equiv \frac{d N}{d (\mu + E_D)} = \frac{2}{\pi}
\frac{k_F}{\hbar v_F + Z(k_F) \partial_{k_F} \Sigma_1}\,,
\label{eqn:TDOS}\\
\hbar \omega_{th} &= E_+(k_F) - E_-(k_F) + \Delta_{e h}\,.
\label{eqn:omega_th}
\end{align}
These formulas contain many-body corrections to the relations given in Sec.~\ref{sec:single} that enter through the derivatives of the self-energy,
while Eq.~\eqref{eqn:omega_th} also has a vertex correction $\Delta_{e h}$.
For example, the DOS $\nu(E)$ [Eq.~\eqref{eqn:nu_E}],
measurable by, e.g., scanning tunneling spectroscopy (STS) is multiplied by the  quasiparticle weight $Z$.
Near the Fermi level one usually finds $Z < 1$ \cite{Giuliani2005qto},
so that the interactions diminish the DOS.
Inferring $v_F$ from $\nu(E_F)$
using the formula $v_F \propto k_F / \nu(E_F)$ of the noninteracting theory
would cause \emph{overestimation} of the Fermi velocity, e.g.,
by the factor of $Z^{-1} = 1 + (1 / 2 + 1 / \pi) \alpha$ at $\alpha \ll 1$ \cite{DasSarma2007mbi}.
[In practice, the low-bias STS data may be influenced by disorder and finite momentum resolution,
see Sec.~\ref{sec:Dirac}.]
Away from the Fermi level the interaction may enhance rather than suppress  $\nu(E)$.
An example is the Dirac point region in a doped SLG
where the DOS is predicted to be nonzero (U-shaped) \cite{LeBlanc2011eoe, Principi2012ttd} rather than vanishing (V-shaped).

Consider next the TDOS $\nu_T(N)$ given by Eq.~\eqref{eqn:TDOS},
which follows from Eqs.~\eqref{eqn:v_q}--\eqref{eqn:nu_E}.
The TDOS can be found by measuring capacitance between graphene and metallic gates,
either stationary \cite{Ponomarenko2010dos, Yu2013ipi} or scanned \cite{Martin2008ooe}.
In the absence of interactions, the TDOS coincides with the DOS at the Fermi level.
However, for repulsive Coulomb interactions the second term in the denominator of
Eq.~\eqref{eqn:TDOS} is negative \cite{Giuliani2005qto}.
(This term can be written in terms of parameter $F_s^0 < 0$ of the Landau Fermi-liquid theory.)
Hence, while $\nu(E_F)$ is suppressed, $\nu_T$ is enhanced compared to the bare DOS.
Extracting $v_F$ from $\nu_T(N)$ \cite{Yu2013ipi}
may lead to \textit{underestimation}.

\begin{figure}[b]
\includegraphics[width=3.20in]{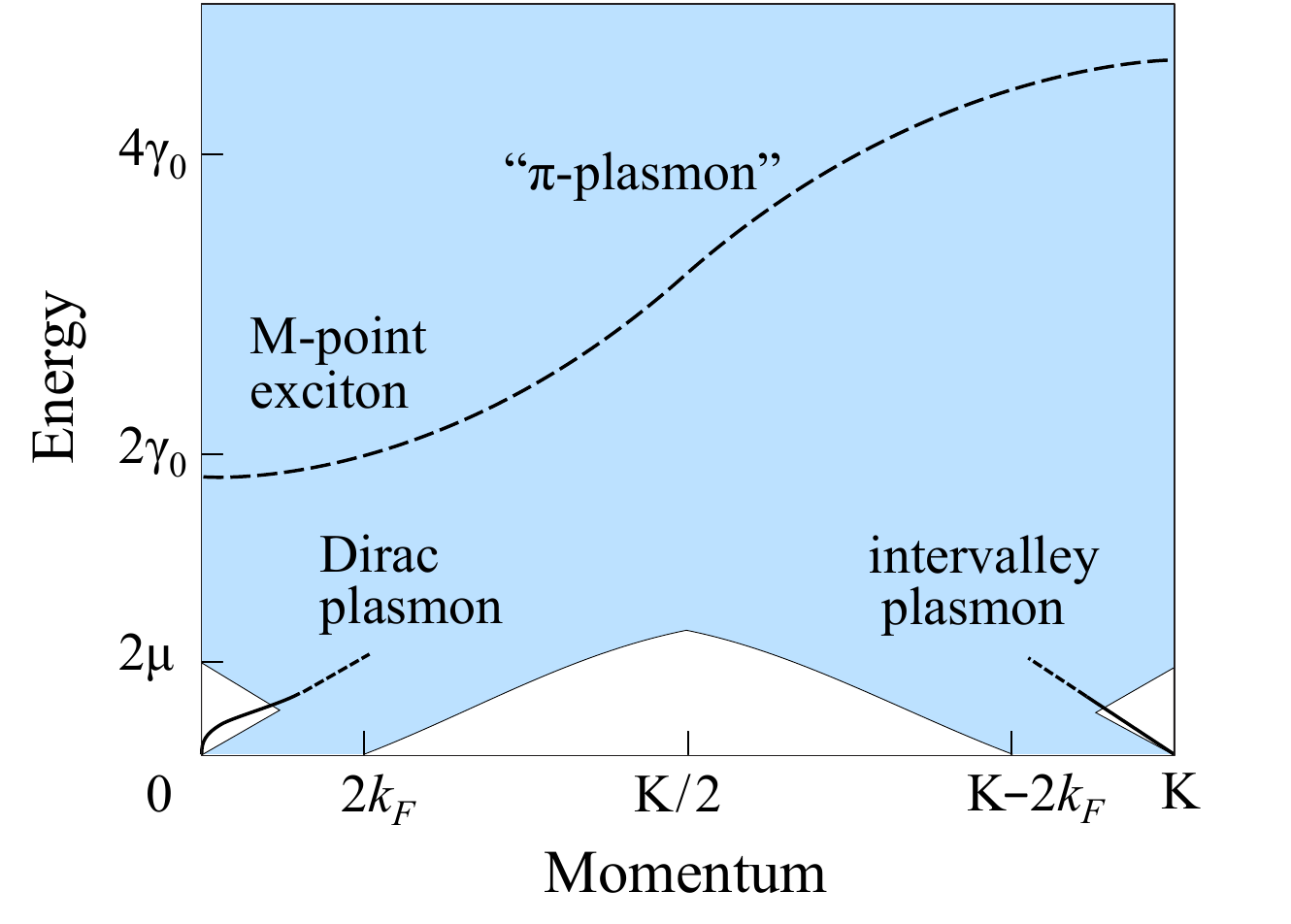}
\caption{\label{fig:Excitations}
Schematic dispersion of electron density excitations in SLG (lines).
The horizontal axis corresponds to the $\Gamma$--$\text{K}$ cut through the Brillouin zone.
All excitations experience Landau damping inside the electron-hole pair continuum (shaded).
}
\end{figure}

The third quantity $\hbar \omega_{th}$ [Eq.~\eqref{eqn:omega_th}]
stands for the threshold energy 
required to excite an electron-hole pair with zero total momentum
in the process of optical absorption.
Without interactions $\hbar \omega_{th} = 2\mu = 2 \hbar v_0 k_F$
(see Fig.~\ref{fig:Excitations}),
and so the bare velocity is equal to $\omega_{th}\, /\, 2 k_F$.
Using the same formula for interacting system \cite{Li2008dcd} may lead
to \emph{underestimation} of the renormalized $v_F$,
for two reasons.
First, $v_F$ is the \emph{group} velocity at the Fermi momentum
while the ratio $[E_+(k_F) - E_-(k_F)] / (2 \hbar k_F)$ gives the average \textit{phase} velocity
of the electron and hole at $q = k_F$.
If the dispersion has the inflection point near $k_F$,
as surmised above, the group velocity must be higher than the phase one.
Second, the threshold energy of the electron-hole pair is reduced by the vertex (or excitonic) correction $\Delta_{e h} < 0$  due to their Coulomb attraction.

Let us now turn to the collective response of SLG at arbitrary $\omega$ and $\mathbf{k}$.
The simplest type of such a process is excitation of a single particle-hole pair
by moving a quasiparticle from an occupied state of momentum $\mathbf{p}$ and energy $E(\mathbf{p}) \leq E_F$ to an empty state of momentum $\mathbf{p}+\mathbf{k} $ and energy $E(\mathbf{p}+\mathbf{k} ) \geq E_F$.
(The subscripts $\pm$ of all $E$'s are again suppressed.)
The particle-hole continuum
that consists of all possible $\bigl(E(\mathbf{p}+\mathbf{k} ) - E(\mathbf{p}), \mathbf{k} \bigr)$ points is sketched in Fig.~\ref{fig:Excitations}.
If the energy and the in-plane momentum of an electromagnetic excitation falls
inside this continuum, it undergoes damping when passing through graphene.
The conductivity $\sigma(\mathbf{k}, \omega) = \sigma' + i \sigma''$ has a finite real part $\sigma'$
in this region.

Collective modes can be viewed as superpositions of many interacting particle-hole excitations.
A number of such modes have been predicted for graphene.
Weakly damped modes exist outside the particle-hole continuum, in the three unshaded regions of
Fig.~\ref{fig:Excitations}.
At low energy the boundaries of these triangular-shaped regions have the slope $\pm \hbar v_F$.
Collective excitations near the $\Gamma$-point (the left unshaded triangle in Fig.~\ref{fig:Excitations}) are Dirac plasmons.
These excitations, reviewed in Sec.~\ref{sec:plasmons},
can be thought of as coherent superpositions of intraband electron-hole pairs from the same valley.
The excitations near the $\mathrm{K}$ point (the right unshaded triangle) involve electrons and holes of different valleys.
Such intervalley plasmons \cite{Tudorovskiy2010ipi} are yet to be seen experimentally.
Also shown in Fig.~\ref{fig:Excitations} are the ``M-point exciton'' that originates from mixing of electron and hole states near the $\mathrm{M}$-points of the BZ (Sec.~\ref{sec:single}) and its finite-momentum extension,
which is sometimes called by a potentially confusing term ``$\pi$-plasmon.''

Two other collective modes have been theoretically predicted but not yet observed and not shown in Fig.~\ref{fig:Excitations}.
One is the excitonic plasmon
\cite{Gangadharaiah2008crf} --- a single interband electron-hole pair marginally bound by Coulomb attraction.
Its dispersion curve is supposed to run near the bottom of the electron-hole continuum.
The other mode \cite{Mikhailov2007nem} is predicted to appear
in the range $1.66 |\mu| < \hbar\omega < 2 |\mu|$ where $\sigma'' < 0$.
Unlike all the previously mentioned collective modes,
which are TM-polarized, this one is TE-polarized.
It is confined to graphene only weakly,
which makes it hardly distinguishable from an electromagnetic wave traveling along graphene.
Besides electron density, collective modes may involve electron spin.
Further discussion of these and of many other interaction effects in graphene can be found in a recent topical review \cite{Kotov2012eei}.

\section{Quasiparticle properties} 
\label{sec:quasiparticles}

\subsection{Dirac spectrum and chirality} 
\label{sec:Dirac}

\begin{figure*}
\includegraphics[width=6.50in]{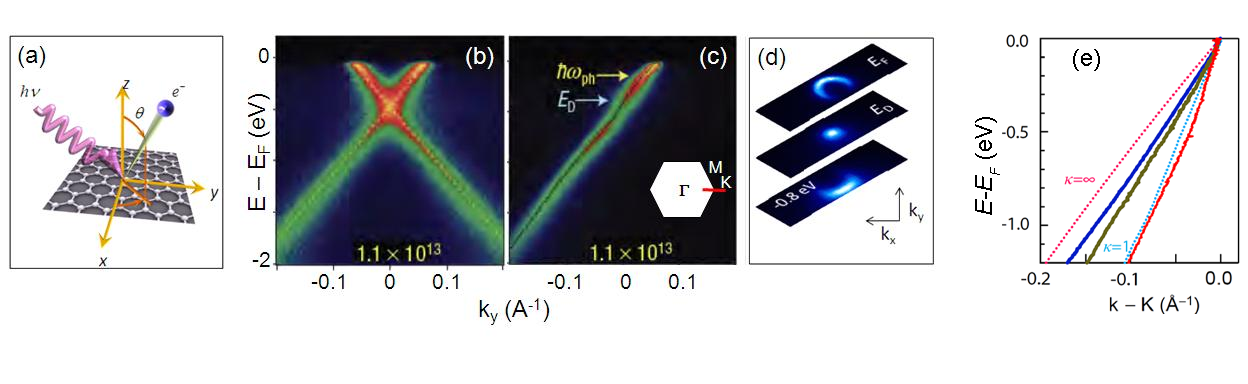}
\caption{\label{fig:ARPES}
(Color online) (a) The ARPES schematics.
(b, c) The ARPES intensity in the energy-momentum space for a potassium-doped epitaxial graphene on SiC (0001) \cite{Bostwick2007qdi}.
In (c) the interval of momenta $k_y$ is indicated by the red line in the inset.
In (b) momentum $k_x$ varies along the orthogonal path through the $\mathrm{K}$-point.
(d) The ARPES maps for a similar sample taken at the energies (top to bottom) $E = E_F \approx E_D +0.4\,\mathrm{eV}$, $E_D$,
and $E_F - 0.8\,\textrm{eV}$ \cite{Hwang2011dmo}.
(e) Solid lines, left to right: the ARPES dispersions along the $\Gamma$--$\mathrm{K}$ direction for graphene on
SiC$(000\bar{1})$, hBN, and quartz.
Dotted lines: results of $\mathrm{GW}$ calculations for $\kappa = \infty$ (magenta) and $\kappa = 1$ (cyan).
Adopted from \cite{Hwang2012fve}.
}
\end{figure*}


The first experimental determination of the SLG quasiparticle spectrum was obtained
by analyzing the Shubnikov-de~Haas oscillations (SdHO) in magnetoresistance \cite{Novoselov2005tda, Zhang2005efm}.
This analysis yields the cyclotron mass
\begin{equation}
m = \hbar k_F / v_F
\label{eqn:m}
\end{equation}
and therefore the Fermi velocity $v_F$.
The lack of dependence of $v_F \approx 1.0 \times 10^8\, \mathrm{cm}/\mathrm{s}$ on the Fermi momentum $k_F$ in those early measurements
was consistent with the linear Dirac spectrum at energies below $0.2\,\mathrm{eV}$.

Direct mapping of the $\pi$-band dispersion over a range of several eV \cite{Zhou2006lee, Bostwick2007qdi} was achieved soon thereafter by the angle-resolved photoemission (ARPES) experiments.
This experimental technique, illustrated by Fig.~\ref{fig:ARPES}(a),
measures the electron spectral function [Eq.~\eqref{eqn:A}] weighted by the
square of the matrix element
$M(\mathbf{k}, \nu)$ of interaction between an incident photon of frequency $\nu$ and an ejected photoelectron of momentum $\mathbf{k}$,
see Eq.~\eqref{eqn:M} below.
The representative dispersion curves measured for epitaxial graphene on SiC are shown in Fig.~\ref{fig:ARPES}(b) and (c),
where red (black) color corresponds to high (low) intensity.
The ``dark corridor'' \cite{Gierz2010idc} $\Gamma$--$\mathrm{K}$
along which one of the two dispersion lines is conspicuously missing, Fig.~\ref{fig:ARPES}(c), occurs due to the selection rules for the matrix element $M(\mathbf{k}, \nu)$ known from prior work on graphite \cite{Shirley1995bzs, Daimon1995sfi}.
The full angular dependence of the ARPES intensity is depicted in Fig.~\ref{fig:ARPES}(d).

The ARPES measurements have been carried out on epitaxial graphene grown on a variety of substrates,
on free-standing samples \cite{Knox2011mar},
and on multilayered samples with weak interlayer interactions \cite{Sprinkle2009fdo}.
The tight-binding model (Sec.~\ref{sec:single}) accounts for the main features of all these spectra.
However, there are also subtle deviations.
For example, the slope of the dispersion near the Dirac point varies systematically with the background dielectric constant $\kappa$ [Fig.~\ref{fig:ARPES}(d)],
which is consistent with the theoretically predicted velocity renormalization,
see Secs.~\ref{sec:interaction} and \ref{sec:renormalization}.
Certain additional features near the Dirac point (see Fig.~\ref{fig:4.1.1})
have been interpreted\footnote{See
\textcite{Zhou2007sib, Gao2010ego,Siegel2012epc,Walter2011eso, Nagashima1994eso, Dedkov2008rei, Varykhalov2008eam, Himpsel1982abd, Enderlein2010tfo, Sutter2009, Rader2009ita, Papagno2012lbg}.} as evidence for substrate-induced energy gaps, Sec.~\ref{sec:substrate}.
For graphene on SiC, an alternative explanation invokes electron-plasmon coupling \textcite{Bostwick2007qdi}, see Fig.~\ref{fig:3.6.2} in Sec.~\ref{sec:e-ph-pl}.

\begin{figure}
\includegraphics[width=3.55in]{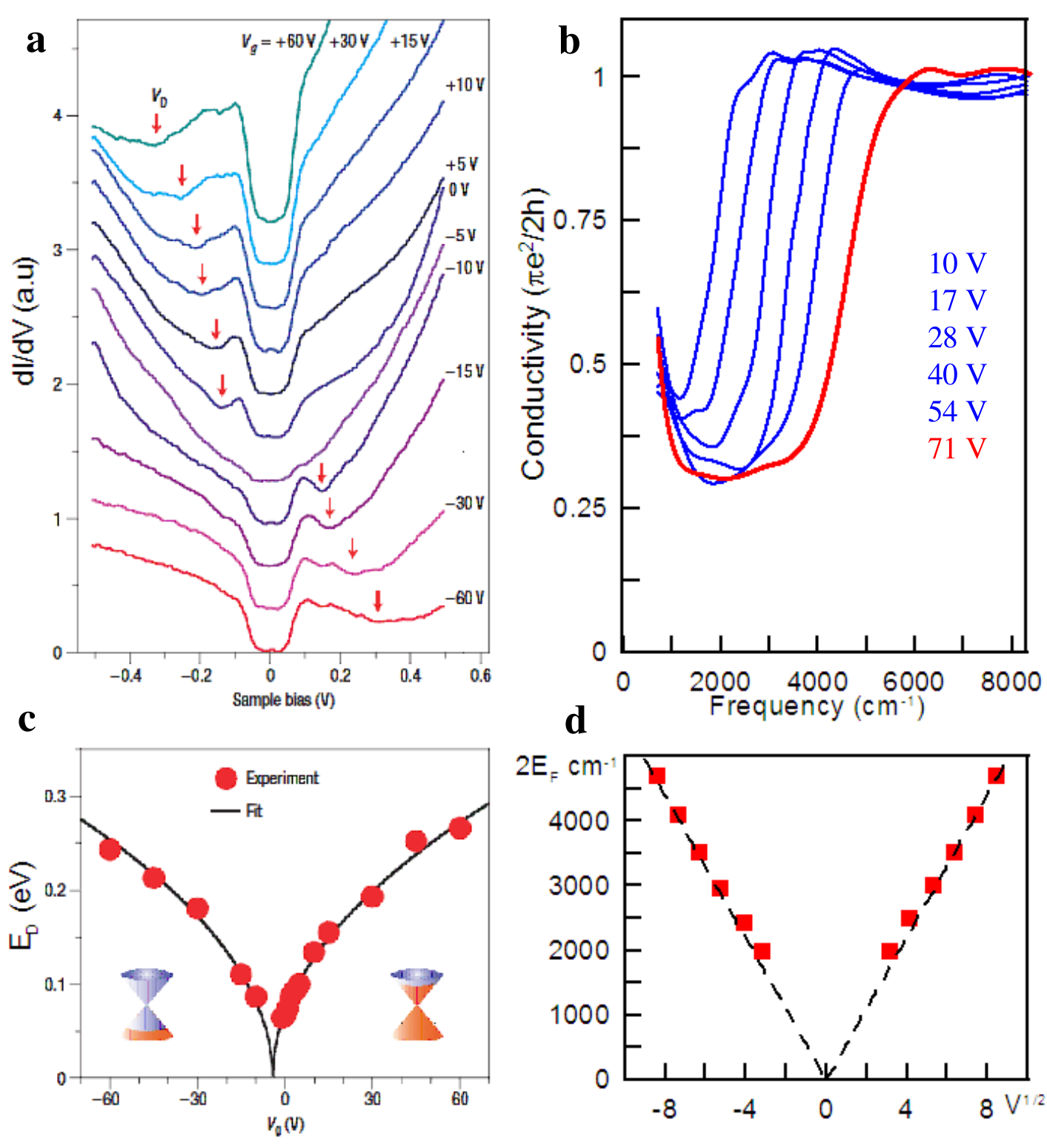}
\caption{\label{fig:3.1.1}
(Color online)
Spectroscopic determination of the Dirac dispersion in SLG.
Panel a: the STS tunneling spectra
$dI/dV$ taken at the same spatial point and different gate voltages $V_g$.
Curves are vertically displaced for clarity.
The arrows indicate the positions of the $dI/dV$ minima $V_D$.
Panel c: the distance $|E_D|$ between the Dirac point and the Fermi level as a function of $V_g$ obtained from the data in panel a.
The line is the fit to $E_D \propto |V_g|^{1/2}$.
The insets are cartoons showing the electron occupation of the Dirac cones. 
After \textcite{Zhang2008gpi}.
Panel b: optical conductivity of SLG at different gate voltages with respect to the neutrality point.
Panel d: the gate voltage dependence of the interband absorption threshold ``$2E_F$''
obtained from the data in panel b.
After \textcite{Li2008dcd}.
}
\end{figure}

Complimentary evidence for the Dirac dispersion of quasiparticles comes the
tunneling and thermodynamic DOS measurements by means of scanned probes.
The Dirac point manifests itself as a local minimum marked by the arrows in the STS tunneling spectra of Fig.~\ref{fig:3.1.1}a.
The U- rather than the V-shaped form of this minimum (Sec.~\ref{sec:single}) is due to disorder smearing.
The STS data obtained by \textcite{Zhang2008gpi} (Fig.~\ref{fig:3.1.1}a) also exhibit
a prominent suppression at zero bias for all gate voltages.
To explain it
\cite{Zhang2008gpi} proposed that this feature arises because of a limitation on the possible momentum transfer in tunneling.
This limitation is lifted via inelastic tunneling accompanied by the emission of a BZ-boundary acoustic phonon of energy $\hbar \omega_0 = 63\,\mathrm{meV}$.
This energy must be subtracted from the tip-sample bias $e V$ to obtain the
tunneling electron energy inside the sample.
By tuning the electron density $N$ with a backgate \cite{Zhang2008gpi, Deshpande2009srs, Brar2010ooc},
one can change the Fermi energy $E_F$ with respect to the Dirac point $E_D$.
Taking the former as the reference point (i.e., assuming $E_F \equiv 0$ for now)
one obtains the relation $|E_D| = |e V_D| - \hbar \omega_0$.
As shown in Fig.~\ref{fig:3.1.1}c,
thus defined $|E_D|$ is proportional to $|N|^{1/2}$, as expected for the linear dispersion,
Eq.~\eqref{eqn:k_F}.
The same zero-bias gap feature is observed in other graphene samples studied by the Berkeley group, e.g., SLG on hBN\cite{Decker2011lepa}.
Yet it is not seen in STS experiments of other groups, see, e.g., Fig.~\ref{fig:el-ph}(c),
Sec.~\ref{sec:renormalization}, and Sec.~\ref{sec:Landau} below.\footnote{See also %
\cite{Deshpande2009srs, Li2012stm, Song2010hrt, Xue2011stm, Yankowitz2012esd, Chae2012rgd}.
}

The $\mu(N)$ dependence can be more directly inferred from the TDOS $\nu_T(N)$ measured by the scanning single-electron transistor microscopy (SSETM) \cite{Martin2008ooe}.
Unlike the STS spectra in Fig.~\ref{fig:3.1.1}a,
the SSETM data are not obscured by the zero-bias feature.
They show a finite and position-dependent TDOS at the neutrality point $N = 0$,
reflecting once again the presence of disorder in graphene on SiO$_2$ substrate, see also Sec.~\ref{sec:Inhomogeneities}.
The most definitive observation of the Dirac TDOS has been made using exfoliated graphene on hBN \cite{Yankowitz2012esd, Yu2013ipi}.
Similar to SSETM, the TDOS was extracted from the capacitance measurements;
however, it was the capacitance between the sample and the global backgate rather than between the sample and the local probe.

Let us now turn to the chirality of graphene quasiparticles.
Recall that chirality refers to the phase relation between the sublattice amplitudes $\psi_j = \psi_j(\mathbf{k} )$, $j = \mathrm{A}, \mathrm{B}$, of the quasiparticle wavefunctions (Sec.~\ref{sec:single}).
The chirality has been independently verified by several techniques.
First, it naturally explains the presence of the special half-filled Landau level at the Dirac point seen in magnetotransport \cite{Novoselov2005tda, Zhang2005efm}.
Next, in the STS experiments the quasiparticle chirality is revealed by the LDOS features observed near impurities and step edges,
see \textcite{Rutter2007sai, Mallet2007eso, Zhang2009oos, Deshpande2009srs} and Sec.~\ref{sec:Inhomogeneities}.
The chirality influences the angular distribution of the quasiparticle scattering by these defects,
suppressing the backscattering \cite{Brihuega2008qci, Xue2012lwl},
in agreement with theoretical predictions \cite{Ando2002dca, Katsnelson2006cta}.

Finally, in ARPES the chirality manifests itself via the selection rules for the matrix element
\begin{equation}
M(\mathbf{k}, \nu) = \frac{e}{c}
\int d \mathbf{r}\,
\Psi_f^*(\mathbf{r}) ({\mathbf{A}} \hat{\mathbf{v}})
\Psi_i(\mathbf{r})
\label{eqn:M}
\end{equation}
that describes coupling of electrons to the vector potential ${\mathbf{A}}$ of the photon.
Here the Coulomb gauge $\bm{\nabla} {\mathbf{A}} = \varphi = 0$ is assumed and $\hat{\mathbf{v}} = -i \hbar \bm{\nabla} / m$ is the velocity operator.
The matrix element $M(\mathbf{k}, \nu )$ depends on 
the relative phase of $\psi_\mathrm{A}$ and $\psi_\mathrm{B}$.
Based on symmetry considerations,
the general form of $M(\mathbf{k}, \nu)$ at small $\mathbf{q} = \mathbf{k} - \mathbf{K}$ must be
\begin{equation}
M(\mathbf{k}, \nu) = (c_1 \mathbf{K} + c_2 \mathbf{q}) \cdot {\mathbf{A}}
\sum_{j = \mathrm{A}, \mathrm{B}} e^{-i \mathbf{K} \bm{\tau}_j}
\psi_j(\mathbf{k} )
\label{eqn:M1}
\end{equation}
if spin-orbit (SO) interaction effects can be ignored.
Here $\bm{\tau}_j$ are the positions of $j$th atom in the unit cell and $\mathbf{K}$ is the nearest Dirac point.
The coefficients $c_1$ and $c_2$ cannot be obtained solely from symmetry;
however, regardless of their values, when $\mathbf{q}$ is parallel (antiparallel) to $\mathbf{K}$ for the states in the conduction (valence) band,
the sum over $j$ in Eq.~\eqref{eqn:M1} vanishes and so does $M(\mathbf{k}, \nu)$.
This explains the low-intensity ``dark corridor'' in the observed ARPES signal, Fig.~\ref{fig:ARPES}(c) and (d).

The ARPES selection rules are also relevant for BLG.
Experimentally, the orientation of the low intensity directions
rotates by $\pm 180^{\circ }$ ($\pm 90^{\circ}$)
in SLG (BLG) when the photon polarization vector ${\mathbf{A}}$ is switched between two orientations, parallel and perpendicular to $\mathbf{K}$ \cite{Gierz2010idc, Hwang2011dmo, Liu2011vec}.
\textcite{Hwang2011dmo} discussed how these rotation angles can be linked to the Berry phase --- a quantity closely related to chirality --- in SLG and BLG\@.
However, their theoretical model for the matrix element $M(\mathbf{k}, \nu)$ has been a subject of controversy,
which appears to be rooted in different assumption about 
the final state wavefunction $\Psi_f(\mathbf{r})$ in Eq.~\eqref{eqn:M}.
At very high energies $h \nu$, the conventional approximation of $\Psi_f(\mathbf{r})$ by a plane wave should be adequate \cite{Shirley1995bzs, Mucha-Kruczynski2008cog}.
In this case one can replace the velocity operator $\hat{\mathbf{v}}$ by $\hbar \mathbf{k} / m$ leading to $c_1 = c_2$ in Eq.~\eqref{eqn:M1}.
On the other hand, \textcite{Hwang2011dmo} replaced $\hat{\mathbf{v}}$ by the band velocity $v\, \mathbf{q} / |\mathbf{q}|$.
This is perhaps appropriate at low energies $h\nu$
at which $\Psi_f \approx \Psi_i$ near the graphene plane.
The corresponding $c_1$ is equal to zero, which is admissible.
However, $c_2 \propto 1 / |\mathbf{q}|$ diverges
at $\mathbf{q} \to 0$, in contradiction to the $\mathbf{k}\cdot \mathbf{p}$ perturbation theory \cite{Yu1996fos}.
In view of this problem and because other ARPES experiments and calculations \textcite{Gierz2010idc}
indicate a nontrivial $\nu$-dependence of $M(\mathbf{k}, \nu)$,
further study of this question is desirable.

\subsection{Renormalization of Dirac spectrum} 
\label{sec:renormalization}

\begin{figure*}
%
\includegraphics[width=7.0in]{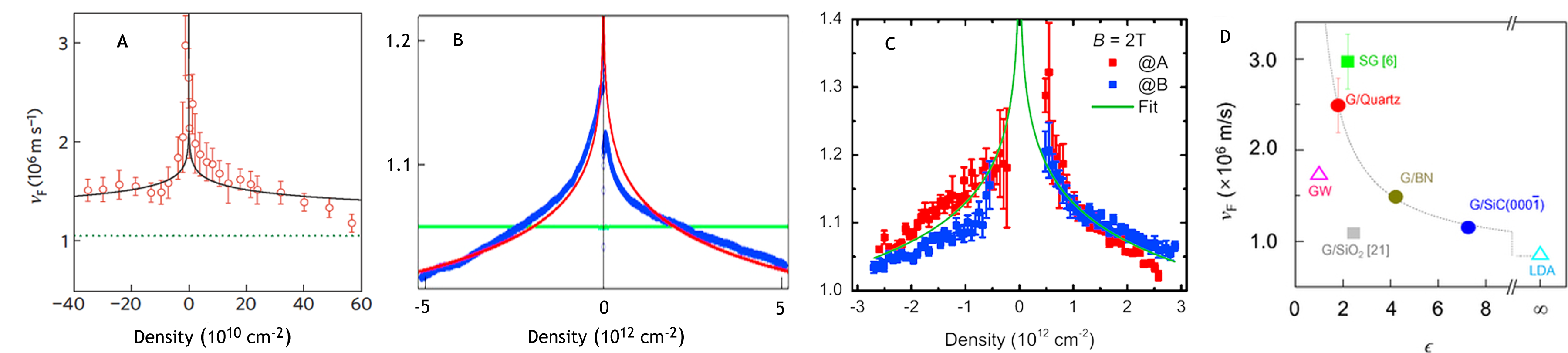}
\caption{\label{fig:v_F}
(Color online)
A. The carrier density dependence of $v_F$ in suspended SLG extracted from magnetoresistance oscillations (circles) and a fit to a theory (solid curve).
Adopted from \cite{Elias2011dcr}.
B. Renormalized velocity determined
from the gate capacitance of SLG on hBN (symbols) and a fit to Eq.~\eqref{eqn:v_k_RG} (solid curve).
After \cite{Yu2013ipi}. 
C. Renormalized velocity of SLG on hBN
from the STS of Landau levels (symbols).
The line is a fit to Eq.~\eqref{eqn:v_F_doped}.
After \cite{Chae2012rgd}.
D. Fermi velocity for SLG as a function of the dielectric constant of the substrate.
The filled symbols are the data points obtained from the ARPES spectra.
The open symbols and the line are from theoretical modeling
\cite{Siegel2012epc}.
%
}
\end{figure*}

Experimental verification of the many-body renormalization of the Dirac spectrum in graphene and its Fermi velocity $v_F$ in particular has been sought after in many spectroscopic studies.
Some of these studies may be subject to interpretation
because $v_F$ usually enters the observables in combination with other quantities, see Sec.~\ref{sec:interaction}.
In addition, 
when the change in $v_F$ is small,
one cannot completely exclude single-particle effects.

Probably the first experimental indication for $v_F$ renormalization in graphene came from infrared absorption/transmission spectroscopy \cite{Li2008dcd} of
exfoliated SLG on amorphous SiO$_2$ (a-SiO$_2$).
This study found that
$v_F$ increases from $1.0 \times 10^8\, \mathrm{cm}/\mathrm{s}$ to
a $15\%$ higher value as the carrier density $N$ decreases from $3.0$ to $0.7 \times 10^{12}\, \mathrm{cm}^{-2}$,
see Fig.~\ref{fig:v_F}d.
Next came an STS study of Landau level spectra \cite{Luican2011qll},
which found a $25\%$ enhancement of $v_F$ (fifth row Table~\ref{tbl:v_F})
in the same range of $N$.

\begin{table}[b]
\centering
\begin{tabular}{l@{\hspace{0.1in}}l@{\hspace{0.2in}}l@{\hspace{0.15in}}l@{\hspace{0.1in}}l}
\hline\hline
Substrate   & $\kappa$  & $v\,(10^8\,\mathrm{cm}/\mathrm{s})$
                                       & Method    & Source\\
\hline
SiC (000\={1}) & 7.26   & 1.15(2)      & ARPES     & Hwang\\
\multirow{2}{*}{hBN}
            &
            \multirow{2}{*}{4.22}
                        & 2.0          & ARPES     & Siegel\\
            &           & 1.20(5)      & Capacitance & Yu\\
SiO$_2$     & 1.80      & 2.5(3)       & ARPES     & Hwang\\
a-SiO$_2$   & 2.45      & 1.47(5)      & STS       & Luican\\
\multirow{2}{*}{Vacuum}
            &
            \multirow{2}{*}{1.00}
                        & 3.0(1)       & SdH       & Elias\\
            &           & 2.6(2)       & Transport & Oksanen\\
\hline\hline
\end{tabular}
\caption{The Fermi velocity of SLG in excess of nominal bare value of $0.85\times 10^8\,\mathrm{cm}/\mathrm{s}$.
In the last column,
Hwang, Siegel, Yu, Luican, Elias, and Oksanen stand for \cite{Hwang2012fve},
\cite{Siegel2013ccs},
\cite{Yu2013ipi}, \cite{Luican2011qll}, \cite{Elias2011dcr},
and \cite{Oksanen2013smm}, respectively.
\label{tbl:v_F}
}
\end{table}

A much broader range of $N$ has been explored in suspended graphene where $N$ as small as a few times $10^{9}\, \mathrm{cm}^{-2}$ can be accessed.
Working with such ultra-clean suspended samples,
\textcite{Elias2011dcr} were able to carry out the analysis of the SdHO
of the magnetoresistance over a two-decade-wide span of the carrier densities.
This analysis yields the cyclotron mass [Eq.~\eqref{eqn:m}]
and thence $v_F$.
The Fermi velocity was shown to reach
$v_F \approx 3.0 \times 10^8\, \mathrm{cm}/\mathrm{s}$,
the largest value reported to date, cf.~Table~\ref{tbl:v_F}.
\textcite{Elias2011dcr} fitted their data (Fig.~\ref{fig:v_F}A) to Eq.~\eqref{eqn:v_k_RG}
for undoped graphene by treating $\alpha$ as an adjustable parameter.
Figure~3 of \textcite{Elias2011dcr}
suggests another possible fit, to Eq.~\eqref{eqn:v_power-law} with the exponent $\beta \approx 0.25$,
which is close to Eq.~\eqref{eqn:beta_N}.
It would be better to compare the measured $v_F$ with
the theoretical predictions for \textit{doped} graphene,
i.e., with the extension (or extrapolation) of Eq.~\eqref{eqn:v_F_doped} to the case in hand, $\alpha \sim 1$.

From the measurements of quantum capacitance (the quantity proportional to the TDOS)
of SLG on hBN,
\textcite{Yu2013ipi} found that $v_F$ increases by $\sim 15\%$ as $N$ varies from $5 \times 10^{12}$ down to a few times $10^{10}\,\mathrm{cm}^{-2}$, see Fig.~\ref{fig:v_F}B.
The vertex corrections were not included when the conversion of the quantum capacitance to $v_F$ was done.
Therefore, this number represents the lower bound on $v_F$, see Sec.~\ref{sec:interaction}.

Using substrates of different dielectric constant $\epsilon_{\mathrm{sub}}$ is another
approach to study $v_F$ renormalization.
An advantage of this approach is that a broad range of $N$ is not necessary in this case.
Instead, the renormalization of velocity is driven by the change in
the interaction strength $\alpha \propto 1 / \kappa$ where $\kappa = (1 + \epsilon_{\mathrm{sub}}) / 2$,
see Eq.~\eqref{eqn:v_k_RG}.
A crude estimate of this effect is as follows.
The dielectric screening by the substrate is effective at distances larger than
the separation $d$ between graphene and the substrate.
Hence, the momentum cutoff in Eqs.~\eqref{eqn:v_k_RG} and \eqref{eqn:v_F_doped}
should be chosen $k_c \sim 1 / d$.
If $d \lesssim 1\,\mathrm{nm}$ and $k_F^{-1} \sim 6\,\mathrm{nm}$,
then $\ln(k_c / k_F) \lesssim 2$ and Eq.~\eqref{eqn:v_k_RG} entails
\begin{equation}
\delta v_F \lesssim (1.0 \times 10^8\, \mathrm{cm}/\mathrm{s}) \times \delta\left(\frac{2}{\epsilon_{\mathrm{sub}} + 1}\right)\,.
\label{eqn:delta_v_F}
\end{equation}
where we use ``$\delta$'' to denote a change in a quantity.
In a recent ARPES study \cite{Hwang2012fve} 
the smallest $v_F = (0.85\pm 0.05) \times 10^{8}\, \text{cm/s}$ was observed on metallic substrates.
This number represents presumably the bare quasiparticle velocity in the absence of long-range Coulomb interactions.
Note that it is close to the Fermi velocity $v_F = 0.81 \times 10^8\,\text{cm}/\text{s}$ measured in carbon nanotubes \cite{Liang2001fpi}.
The ARPES results for three other substrates are reproduced
in Fig.~\ref{fig:v_F}D.
They clearly demonstrate a prominent velocity enhancement near the Fermi level.
Thus, graphene on (the carbon face of) SiC has $v_F$ that is only slightly larger than what is observed for metallic substrates \cite{Siegel2011mbi, Hwang2012fve},
which can be explained by the high $\kappa$.
Graphene on hBN has $v_F$ close to that for SLG on a-SiO$_2$,
which is consistent with the effective dielectric constants of hBN and a-SiO$_2$ being roughly equal~\cite{Wang2012mdq, Yu2013ipi}.
A surprisingly large $v_F$ is found for graphene on crystalline SiO$_2$ (quartz),
see Table~\ref{tbl:v_F} and Fig.~\ref{fig:v_F}D.

\begin{figure*}
\includegraphics[width=6.50in]{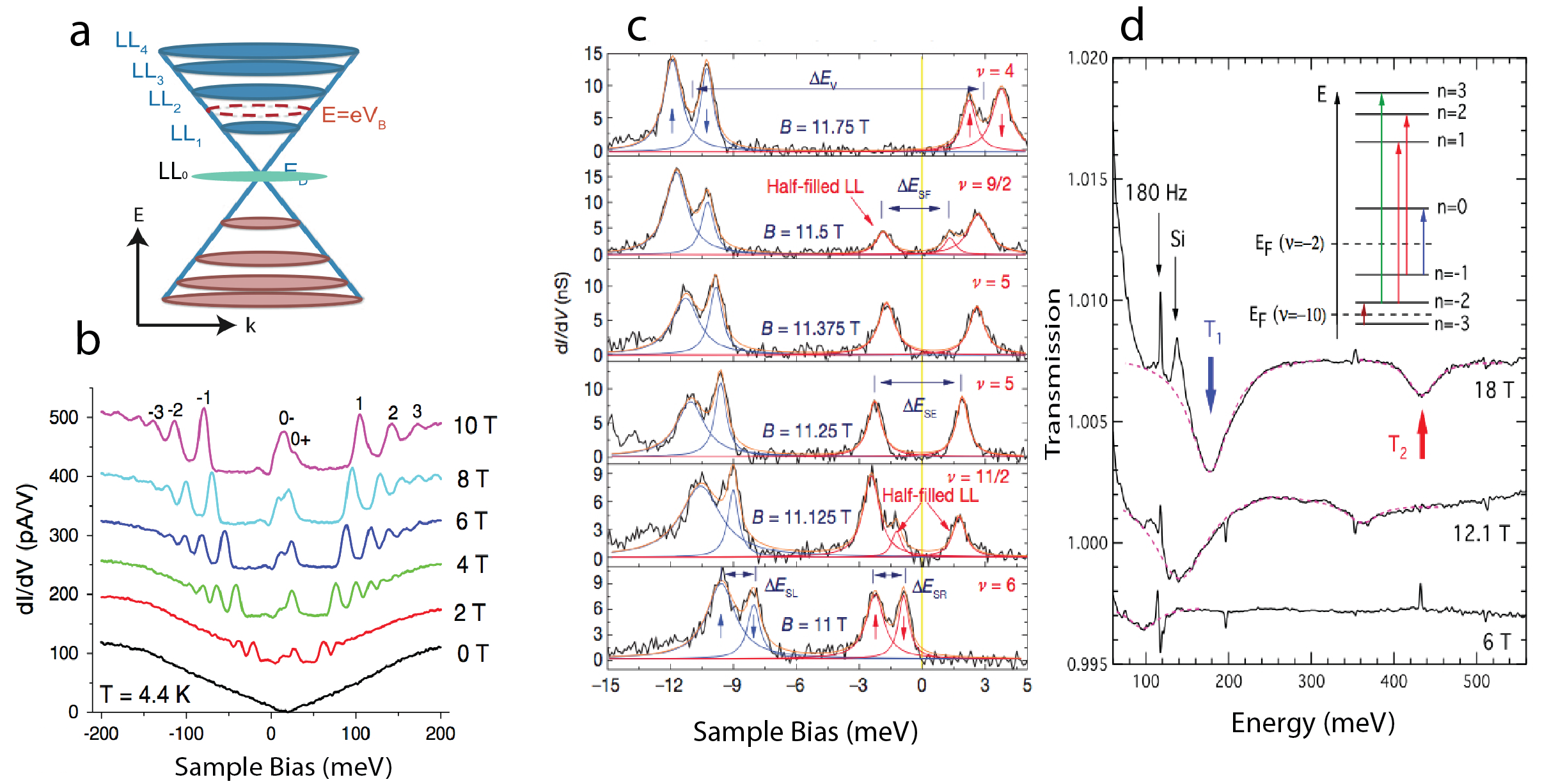}
\caption{\label{fig:Landau_levels}
(Color online)
(a) The schematics of LLs in SLG.
Each LL is four-fold degenerate due to spin and valley degrees of freedom.
The neutrality point corresponds to the half-filling of the $n=0$ LL.
(b) The STS spectra of graphene on graphite at different magnetic fields \cite{Li2009sts}.
(c) A high-resolution STS revealing four-fold states that make up $n=1$ LL of epitaxial graphene on SiC at different magnetic fields.
The energy separations $\Delta E_{v}$ and $\Delta E_{s}$ due to lifting of the valley and spin degeneracies are enhanced when the Fermi level falls between the spin-split levels at filling factor $\nu =5$.
Additional stable states appear at $\nu = 11/2$ and $9/2$ \cite{Song2010hrt}.
(d) The IR transmission of a $p$-doped graphene at $\nu = -2$ normalized to that at $\nu = 10$ at three different magnetic fields.
Two LL resonances, $T_{1} $ and $T_{2} $, are observed.
The inset shows the allowed LL transitions
\cite{Jiang2007iso}.}
\end{figure*}

As mentioned above, renormalization of the quasiparticle velocity in SLG can also arise from single-particle physics.
One example is the modification of the electron band-structure by external periodic potentials \cite{Park2008ngo, Park2008abo, Brey2009ezm, Guinea2010bsa, Wallbank2013gms}.
Such potentials are realized in moir\'e superlattices that form when graphene is deposited on lattice-matched substrates,
which we will discuss in Sec.~\ref{sec:moire}.
Similar effects appear in misoriented graphene bilayers and multilayers
that grow on the carbon face of SiC \cite{Haas2008wmg} (Sec.~\ref{sec:forms}) and are also common in CVD graphene grown on Ni \cite{Luican2011slb}.
Calculations predict a strong dependence of the velocity on the twist angle \cite{Bistritzer2011mbi, TramblydeLaissardiere2010lod, LopesdosSantos2007gbw, Shallcross2010eso,
Lopes_dos_Santos2012cmo}.
%
The experimental value of $v_{F}$ reported for twisted graphene layers on the carbon face of SiC is $v_{F } \approx 1.10 \times 10^{8}\,\text{cm/s}$ \cite{Crassee2011mmo, Miller2009otq, Siegel2011mbi, Sprinkle2009fdo}.
Changes of $v_{F}$ up to 10\% among different layers for graphene on the carbon-face of SiC have been deduced from SdH oscillations \cite{deHeer2007eg} and magneto-optical measurements \cite{Crassee2011mmo, Crassee2011gfr}.
In the latter case these changes have been attributed to electron-hole asymmetry and also to variation of the carrier density and dielectric screening among the graphene layers.
No variation of $v_{F}$ as a function of twist angle was observed by ARPES and STS \cite{Miller2009otq, Sprinkle2009fdo, Siegel2011mbi, Sadowski2006lls}.
However,
a $14\%$ decrease of $v_{F}$ at small twist angles was found in the STS study of CVD graphene transferred to the grid of a transmission electron microscope \cite{Luican2011slb}.

\subsection{Landau quantization} 
\label{sec:Landau}

Spectroscopy of Landau level (LL) quantization in a magnetic field is yet another way to probe quasiparticle properties of graphene.
The linear dispersion of SLG leads to unequally spaced LLs: $E_{n} = E_D + \text{sgn}(n) v_0 \sqrt{2 e \hbar B |n|} $ (Fig.~\ref{fig:Landau_levels}a), where $n > 0$ or $n < 0$ represents electrons or holes, respectively\cite{McClure1957bso, Gusynin2006tod, Jiang2007iso}.
Each of the LLs has four-fold degeneracy due to the spin and valley degrees of freedom.
Additionally, the electron-hole symmetric $n = 0$ LL
gives rise to the extraordinary ``half-integer'' quantum Hall effect \cite{Novoselov2005tdg, Zhang2005eoo},
the observation of which back in 2005 was the watershed event that ignited the widespread interest in graphene.

The LL spectrum of graphene has been probed using scanning tunneling spectroscopy (STS), IR spectroscopy, and Raman scattering.
The STS of graphene LLs was first carried out in graphene on graphite samples, where suspended graphene is isolated from the substrate at macroscopic ridge-like defect in graphite \cite{Li2009sts}.
Figure~\ref{fig:Landau_levels}b displays the differential conductance of graphene versus tip-sample bias at different magnetic fields $B$ normal to the graphene surface.
Well defined LDOS peaks corresponding to discrete LL states appear in the tunneling spectra.
These LL peaks become more prominent and shift to higher energies in higher magnetic fields
consistent with the expected $\sqrt{B |n|}$ law.
Similar LL spectrum was also observed in epitaxial grown graphene layers on SiC \cite{Miller2009otq}.

To examine the fine structure within a LL,
\textcite{Song2010hrt} performed high-resolution STS studies at temperatures as low as $10\,\text{mK}$ on epitaxial graphene.
Figure~\ref{fig:Landau_levels}c shows their data for the $n = 1$ LL at the magnetic field range where the LL1 starts to cross the Fermi energy (yellow line).
The LL1 level is composed of four separate peaks, indicating that the valley and spin degeneracy is lifted.
The larger energy splitting ($\Delta E_{v}$) is attributed to the lifting of valley degeneracy.
It increases monotonically with the applied magnetic field with the effective $g$-factor of $18.4$.
The smaller splitting ($\Delta E_{s} $) has an average $g$-factor close to $2$, presumably due to the electron spin.
Quantitatively, this spin splitting shows a highly unusual dependence on the filling factor.
Comparing the spectra at filling factors of $4$, $5$, and $6$,
a clear enhancement of the spin splitting is observed at $\nu =5$,
which can be attributed to many-body effects (exchange enhancement).
In addition, new stable half-filled Landau levels appear at half fillings such as $9/2$ and $11/2$.
Their origin is not yet clear.
Landau level spectroscopy of graphene on SiO$_2$ was presented in \textcite{Luican2011qll} and a similar study for graphene on hBN was reported in \textcite{Chae2012rgd}.
In the latter system, which has lower disorder, observation of many LLs was possible over a wide energy range.
Deviations of the LL energies by about $\sim 10\%$ from the predictions of the single-particle theory were interpreted in terms of the Fermi velocity renormalization, see Fig.~\ref{fig:v_F}C.
This is in line with the results of other measurements discussed above (Table~\ref{tbl:v_F}).

The infrared (IR) spectroscopy provides another way to study the LL spectra \cite{Sadowski2006lls, Jiang2007iso, Henriksen2010iis}.
The IR transitions between LLs have to satisfy the selection rule
$\Delta |n|=\pm 1$, due to angular momentum conservation.
Selection rules also apply to the circular polarization of light.
As a result, graphene exhibits strong circular dichroism and Faraday effect  \cite{Crassee2011gfr}.
Figure~\ref{fig:Landau_levels}d displays the experimental data of normalized IR transmission spectra through SLG at several magnetic fields \textcite{Jiang2007iso}.
The electron density is controlled so that Fermi energy lies between the $n = -1$ and $0$ LL (inset in Fig.~\ref{fig:Landau_levels}d).
Two transmission minima $T_{1} $ and $T_{2}$ are readily observable.
The $T_{1} $ resonance corresponds to the $n=-1$ to $n=0$ intraband LL transition, and the $T_{2} $ resonance arises from the degenerate interband $n=-1$ to $n=2$ and $n=-2$ to $n=1$ transitions.
The LL transition energies scales linearly with $\sqrt{B}$, as expected from the LL structure described above.
A careful examination of the IR transitions as a function of electron filling factor further reveals
that at zero filling factor, the $n=-1$ to $n=0$ (or $n=0$ to $n=1$) transition is shifted to a higher energy compared to that at the filling factor of $2$ and $-2$ \cite{Henriksen2010iis}.
This shift was again tentatively attributed to interaction effects.

\section{Current/density response and collective modes} 
\label{sec:Excitations}

\subsection{Optical conductivity}
\label{sec:direct}

Traditionally measured by optical spectroscopy,
the ``optical'' conductivity $\sigma(\omega) = \sigma'(\omega) + i \sigma''(\omega) \equiv \sigma(q = 0, \omega)$ quantifies the
response of current to an external electric field in the low momenta $q \ll \omega / v_F$ region of the $q$-$\omega$ parameter space,
see Fig.~\ref{fig:Excitations}.
Both intraband and interband transitions contribute to the optical conductivity;
we will start with the interband ones.

In a charge-neutral SLG, which is a zero-gap semiconductor with the Fermi energy at the Dirac point,
the interband transitions have no threshold.
Particularly interesting is the range of (IR) frequencies $\hbar \omega \ll \gamma_0$,
where quasiparticles behave as massless Dirac fermions.
Since the Dirac spectrum has no characteristic frequency scale and neither does the Coulomb interaction,
at zero temperature and in the absence of disorder the conductivity must be of the form
$\sigma(\omega) = (e^2 / h) f(\alpha)$, where $\alpha$ is defined by Eq.~\eqref{eqn:alpha}.
[However, $\omega = 0$ is, strictly speaking, a singular point \cite{Ziegler2007mco}.]
For the noninteracting case, $\alpha = 0$, the theory predicts\footnote{%
See \textcite{Ludwig1994iqh} for Dirac fermions in general and
\textcite{Ando2002dca, Gusynin2006tod, Peres2006epo, Ziegler2007mco, Falkovsky2007std, Stauber2008oco} for SLG\@.}
$f(0) = \pi / 2$, so that $\sigma(\omega)$ is real and has the universal value of
\begin{equation}
\sigma_0 = \frac{\pi}{2}\, \frac{e^2}{h}\,.
\label{eqn:sigma_0}
\end{equation}
The corresponding transmission coefficient $T = 1 - 4\pi \sigma(\omega) / c$
for suspended graphene is expressed solely in terms of the fine structure constant:
$T = 1 - \pi (e^2 / \hbar c) \approx 0.977$.\footnote{See \textcite{Abergel2007vog, Blake2007mgv, Roddaro2007tov, Ni2007gtd}.}
This prediction matches experimental data surprisingly well,
with possible deviations not exceeding $15\%$ throughout the IR and visible spectral region
\cite{Mak2008mot, Nair2008fsc, Li2008dcd}.
This implies that the interaction correction $f(\alpha) - f(0)$ is numerically small
even at $\alpha = 2.3$.
At the level of the first-order perturbation theory this remarkable fact is explained by
a nearly complete cancellations between self-energy and vertex contributions \cite{Mishchenko2008mci, Sheehy2009oto, Sodemann2012ict}

Doping of graphene creates an effective threshold $\hbar \omega_{th}$ for interband absorption by the same mechanism as in the Burstein–Moss effect:
the blue shift of the lowest energy of interband transitions in a doped semiconductor~\cite{Yu1996fos}.
Due to Pauli blocking,
no direct interband transitions exist at $\hbar\omega < 2 |\mu|$
in the noninteracting electron picture, see Fig.~\ref{fig:Excitations}.
Experimentally, the existence of such a threshold has been confirmed 
by IR spectroscopy of gated SLG \cite{Wang2008gvo, Li2008dcd, Horng2011dco}.
As shown in Fig.~\ref{fig:3.1.1}, the frequency position of the broadened step in $\sigma'(\omega)$ scales as
the square-root of the gate voltage $V_g$, and so is proportional to $k_F$.
This is consistent with the linear dispersion $2 |\mu| = 2 \hbar v_F k_F$ of the Dirac quasiparticles.
This behavior is seen in both exfoliated \cite{Li2008dcd} and CVD-grown graphene \cite{Horng2011dco}.
At the smallest gate voltages,
deviations from the square-root law are seen \textcite{Li2008dcd},
which may be due to an interplay of many-body effects,
the velocity renormalization and the vertex corrections,
see Secs.~\ref{sec:interaction} and \ref{sec:renormalization}.

Vertex corrections (which are also referred to as the excitonic effects) play a prominent role also in the optical energy range $4$--$6\,\text{eV}$.
The dominant spectroscopic feature in this region is the interband transition that connects electron and hole states near the $\mathrm{M}$-point of the Brillouin zone,
where the DOS has van~Hove singularities (Sec.~\ref{sec:single}).
This resonance is seen both in SLG and MLG samples.\footnote{See \textcite{Fei2008heo, Santoso2011oor, Mak2011smb, Kravets2010seo, Chae2011efr}.}
This resonance has been detected by EELS and dubbed ``$\pi$-plasmon'' [see, e.g., \textcite{Eberlein2008pso}].
We prefer the term ``$\mathrm{M}$-point exciton'' to avoid confusion with the Dirac plasmon.
Electron-electron interactions significantly renormalize the properties of this resonance.
The position of the $\mathrm{M}$-point exciton is red shifted from the noninteracting value of $2\gamma_0$
by as much as $600\,\text{meV}$ in SLG samples \cite{Yang2009eeo, Mak2011smb, Chae2011efr}.
The absorption peak has a Fano lineshape indicative of interaction effects.

Let us now discuss the intraband transitions.
The commonly used Drude model assumes that the intraband response of a conductor is a simple fraction:
\begin{equation}
\label{eqn:sigma_Drude}
\sigma_{\mathrm{intra}}(\omega) = \frac{i}{\pi}\, \frac{D}{\omega + i\gamma}\,,
\end{equation}
For noninteracting electrons with an isotropic Fermi surface one generally finds \cite{Ashcroft1976ssp}
\begin{equation}
D = \pi e^2 |N| / m\,,
\label{eqn:D}
\end{equation} 
where $m$ is defined by Eq.~\eqref{eqn:m}.
For Dirac electrons with $v_F = \mathrm{const}$ and $k_F = \sqrt{\pi |N|}$ both $m$ and $D$ scale as $|N|^{1/2}$.
Parameter $D$ is known as the Drude weight.
In the Drude model, the relaxation rate $\gamma$ is frequency-independent and can be related to the transport mobility $\mu_{tr}$ by $\hbar \gamma = e v_F / (k_F \mu_{tr})$.
In exfoliated samples of typical mobility $\mu_{tr} \sim 10,000\,\text{cm}^{2}/ \text{Vs}$ and carrier density $N \sim 3\times 10^{11}\,\mathrm{cm}^{-2}$ one estimates $\gamma \sim 10\,\mathrm{meV}$.
This is below the low-frequency cutoff of the IR microscopy \cite{Li2008dcd}.
One can extend measurements to lower frequency provided larger area samples are used,
such as epitaxial \cite{Choi2009ber, Hofmann2011hcc} and CVD-grown graphene \cite{Horng2011dco, Rouhi2012tgo, Ren2012tai}.
In both cases the gross features of the measured frequency dependence of IR conductivity comply with the Drude model.
Note that such samples have relatively low mobility (Sec.~\ref{sec:forms}) and so show wider Drude peaks
in $\sigma'(\omega)$.

\begin{figure}
\includegraphics[width=2.2in]{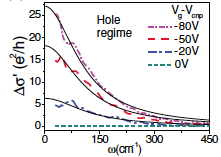}
\caption{\label{fig:3.2.1}
(Color online) Gating-induced change
$\Delta \sigma'(\omega) = \sigma'(\omega) - \sigma'_{\mathrm{CNP}}(\omega)$
in the optical conductivity of SLG.
Solid lines are the fits assuming Drude model for both
$\sigma(\omega)$ and $\sigma_{\mathrm{CNP}}(\omega)$.
The latter is the conductivity at the charge-neutrality point.
Its Drude form is chosen to account for inhomogeneous local doping, cf.~Sec.~\ref{sec:Inhomogeneities}.
After \cite{Horng2011dco}.}
\end{figure}

The intraband response as a function of the carrier density
has been studied using a gated CVD-grown graphene \cite{Horng2011dco}.
The experimentally observed Drude weight was found to be 20--50\% smaller than predicted by
Eq.~\eqref{eqn:D}, see Fig.~\ref{fig:3.2.1}.
The reduction was larger on the electron ($\mu > 0$) side where the transport mobility was also lower.
At the same time, the optical sum rule $\int_0^\infty \sigma'(\omega) d \omega = \mathrm{const}$
was apparently obeyed \cite{Horng2011dco}.
The conservation of the total optical weight was made possible
by a residual conductivity 
in the interval $\gamma \ll \omega \ll 2 |\mu| - \gamma$,
first observed by \textcite{Li2008dcd}.
In this region of frequencies both interband and intraband
transitions should be suppressed
yet the conductivity remains no smaller than $\sigma'(\omega) \approx 0.5 e^2 / h$, see Fig.~\ref{fig:3.1.1}b.
Redistribution of the optical weight is common to correlated electron systems \cite{Millis2004oca, Qazilbash2009eci, Basov2011eoc}, and so the residual conductivity of graphene is suggestive of interaction effects.
Calculation of such effects is more difficult than for the undoped graphene but an extensive theoretical literature already exists on the subject.
For example, the role of interaction in the conductivity sum rule was tackled in \cite{Sabio2008fsr},
the renormalization of $D$ was discussed in \cite{Abedpour2007rou, Levitov2013eei}.
The residual conductivity remains the most challenging problem.
So far, theoretical calculations that consider electron-phonon\footnote{%
See \textcite{Stauber2008cos, Peres2008tic, Hwang2012ose, Scharf2013eos}.}
or electron-electron\footnote{%
See \textcite{Grushin2009eoc, Peres2010eei, Hwang2012ose, Carbotte2012eop, Principi2013ild}.}
interactions predict relatively small corrections to $\sigma'(\omega)$ inside the interbad-intraband gap $0 < \hbar \omega < 2 |\mu|$.
Such corrections can however be enhanced by disorder \cite{Kechedzhi2013pad, Principi2013idd}.

\subsection{Plasmons} 
\label{sec:plasmons}

%
%
\begin{figure*}
\includegraphics[width=6.0in]{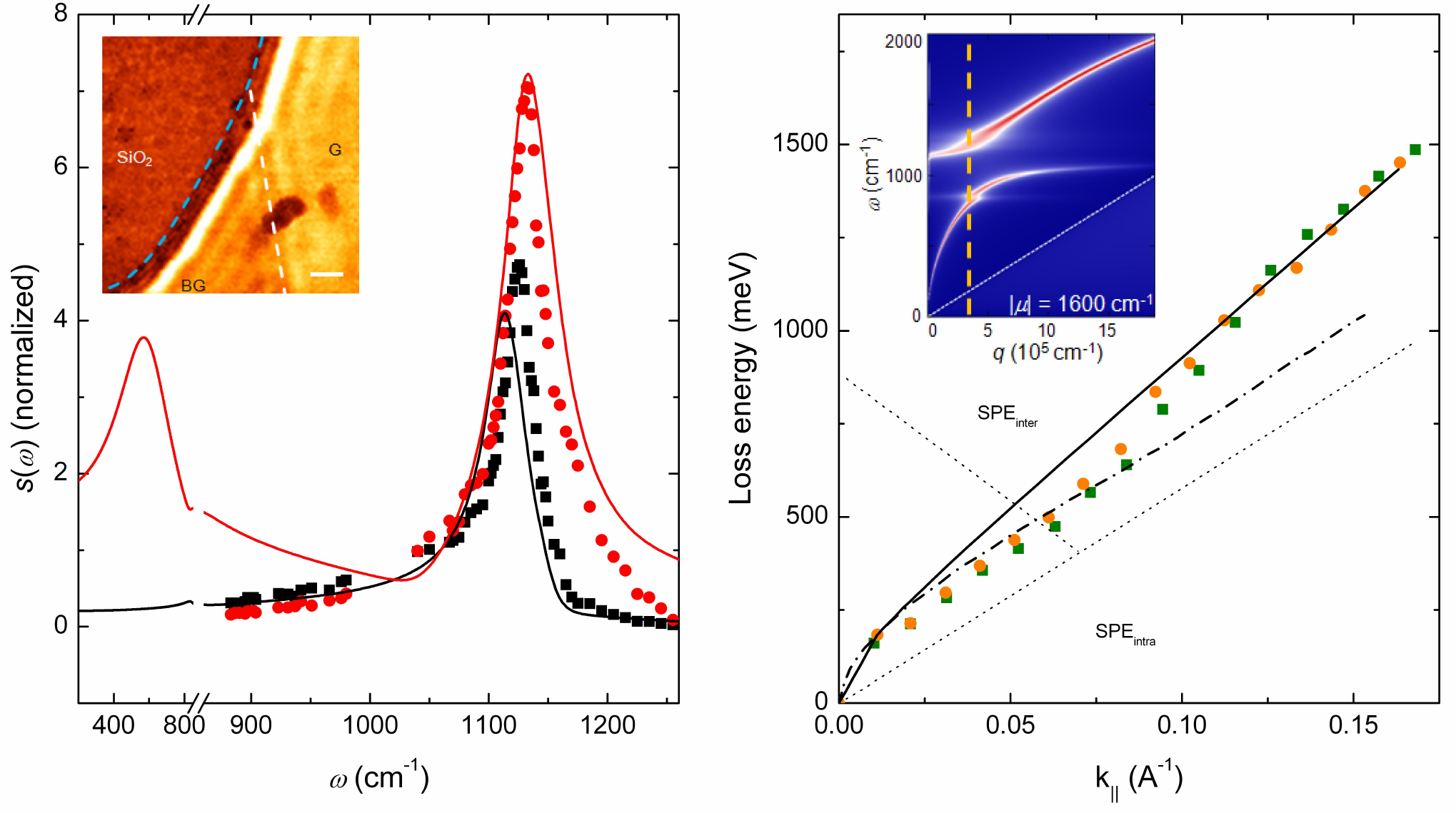}
\caption{
\label{fig:3.3.1}
(Color online) Collective modes of graphene on polar substrates originate from hybridization of substrate surface phonons with graphene plasmons.
Both modes show up as resonances in the near-field amplitude spectrum $s(\omega)$.
The main left panel shows the phonon mode measured for SiO$_2$ alone (black squares) and
the phonon-plasmon hybrid mode of SiO$_2$ covered with SLG (red dots).
The modeling results are shown by the lines,
with the SLG trace revealing the lower hybrid mode of predominantly plasmon character at $\omega \sim 500\,\text{cm}^{-1}$.
[After \textcite{Fei2011ino}.]
Direct observation of this plasmon-like mode is achieved by real-space imaging of $s(\omega)$ at a fixed frequency
$\omega = 892\,\text{cm}^{-1}$ (the inset).
The oscillations seen in the image result from interference of plasmon waves \cite{Fei2012gto}.
The bright lines in the right inset depict the calculated mode dispersions
for SLG with the chemical potential $\mu / h c = 1600\,\text{cm}^{-1}$ on SiO$_2$.
The experimentally relevant momenta are situated near the vertical dashed line.
The diagonal dashed line is the border of the electron-hole continuum (cf.~Sec.~\ref{sec:interaction}).
The main right panel depicts collective modes of epitaxial graphene on SiC measured with electron energy loss spectroscopy at $300\,\text{K}$ (dots) and $80\,\text{K}$ (squares).
The solid and the dash-dotted lines are different theoretical fits.
The dotted lines indicate the boundaries of the electron-hole continuum [from~\textcite{Tegenkamp2011peh}].
}
\end{figure*}

A plasmon is a collective mode of charge-density oscillation in a system with itinerant charge carriers.
Plasmons have been extensively investigated both in classical and quantum plasmas.
The dispersion relation of plasmons in a 2D conductor is given by the  equation
\begin{equation} \label{eqn:q_p}
q_p(\omega) = \frac{i}{2\pi}\,
  \frac{\kappa(\omega) \omega}{\sigma(q_p, \omega)}\,,
\end{equation} 
where $\kappa(\omega)$ is the average of the dielectric functions of the media on the two sides, see, e.g., \cite{Fei2012gto, Grigorenko2012gp}.
At $q \ll k_F$ the $q$-dependence of $\sigma(q, \omega)$ can be neglected,
and so the plasmon dispersion is determined by the optical conductivity
$\sigma(\omega)$ discussed above.
This implies that $\sigma(\omega)$, which is usually measured by optical spectroscopy,
can also be inferred by studying plasmons \cite{Fei2012gto, Chen2012oni}.
(Actually, optics probes transverse rather than longitudinal response but at $q \ll\omega / v_F$ the two coincide.)

Note that $q_p = q_p' + i q_p''$ is a complex number.
Its real part determines the plasmon wavelength $\lambda_p = 2\pi / q_p'$ and the imaginary part characterizes dissipation.
The condition for the propagating plasmon mode to exist is $q_p'' \ll q_p'$ or $\sigma' \ll \sigma''$,
assuming $\kappa$ is real.
In SLG this condition is satisfied (both in theory and in experiment)
at frequencies that are smaller or comparable to $|\mu| / \hbar$.
In particular,
at $\hbar \omega \ll |\mu|$, one can use Eqs.~\eqref{eqn:sigma_Drude} and \eqref{eqn:q_p}
to express the plasmon dispersion in terms of the Drude weight $D$:
\begin{equation} \label{eqn:omega_I}
\omega_p(q) = \sqrt{\frac{2}{\kappa}\, D q}\,.
\end{equation} 
This $\sqrt{q}$-behavior is a well-known property of 2D plasmons.
Using \eqref{eqn:D} for $D$ with Eq.~\eqref{eqn:m} for $m$,
one finds
\begin{equation} \label{eqn:omega_p_q} 
\omega_p(q) = \sqrt{\frac{2 \sqrt{\pi}\,e^2}{\kappa\hbar v_F}}\ v_F |N|^{1 /4} q^{1/2}\,,
\quad q \ll k_F\,.
\end{equation} 
The $|N|^{1/4}$-scaling of the plasmon frequency at fixed $q$ should be contrasted with $\omega_p \propto |N|^{1/2}$ scaling well known for the 2D electron gas with a parabolic energy spectrum (2DEG).
The difference is due to the $D \propto N$ dependence in the latter system.
Another qualitative difference is the effect of electron interactions on $D$.
In 2DEG, interactions do not change $D$,
which is the statement of Kohn's theorem~\cite{Giuliani2005qto}.
In graphene, interactions renormalize the Drude weight~\cite{Abedpour2007rou, Levitov2013eei},
which causes quantitative deviations from Eq.~\eqref{eqn:omega_p_q}.
Qualitative deviations from this equation occur however only at $q \sim  k_F$ where
the plasmon dispersion curve enters the particle-hole continuum,
see Fig.~\ref{fig:Excitations}.
At such momenta the Drude model~\eqref{eqn:sigma_Drude} 
breaks down
and a microscopic approach such as the random-phase approximation (RPA) becomes necessary \cite{Wunsch2006dpo, Hwang2007dfs, Jablan2009pig}.
The RPA predicts that  inside the particle-hole continuum the plasmon survives as a broad resonance that disperses
with velocity that approaches a constant value $v_F$ at large $q$.

Experimental measurements of the plasmon dispersion over a broad range of $q$ have been obtained by means of electron energy loss spectroscopy (EELS).
Such experiments\footnote{See \textcite{Liu2008pda, Liu2010pps, Koch2010spp, Shin2011cot, Tegenkamp2011peh}.} have confirmed the $\omega_p \propto \sqrt{q}$ scaling at small momenta and a kink in the dispersion in the vicinity of the particle-hole continnum.
EELS study carried out by \textcite{Pfnur2011mpe} reported two distinct plasmon modes,
a result yet to be verified through other observations.

The IR spectroscopy of graphene ribbons \cite{Ju2011gpf, Yan2013dpm} and disks \cite{Yan2012ist, Yan2012pcg, Fang2013gth}
offered a complementary method to probe the plasmon dispersion.
The experimental signature of the plasmon mode is the absorption resonance whose frequency $\omega_{\mathrm{res}}$ is observed to scale as the inverse square root of the ribbon width $W$ (or disk radius $R$).
This scaling agrees with the theoretical results relating $\omega_{\mathrm{res}}$ to the plasmon dispersion in an unbounded graphene sheet [Eq.~\eqref{eqn:omega_I}].
For the ribbon, it reads $\omega_{\mathrm{res}} \approx \omega_p(2.3 / W)$ \cite{Nikitin2011ewt}.
The same relation can be deduced from the previous work \cite{Eliasson1986mms} on plasmons in 2DEG stripes.
In fact, most of the results obtained in the context of plasmons in 2DEG in semiconductors \cite{Demel990ndr, Demel1991odp, Kukushkin2003ore} and also electrons on the surface of a liquid $^4$He \cite{Glattli1985dhe, Mast1985obe} directly apply to graphene whenever the Drude model holds.

As shown theoretically and experimentally in that earlier work,
the spectrum of plasmons in ribbons/stripes is split into a set of discrete modes dispersing as $\omega_l(q_\parallel) \approx \omega_p\bigl(\sqrt{q_l^2 + q_\parallel^2}\,\bigr)$,
as a function of the longitudinal momentum $q_\parallel$
and the mode number $l = 1, 2, \ldots$,
with $q_l = (\pi l - \delta_l) / W$ having the meaning of the transverse momentum.
Numerical results \cite{Eliasson1986mms, Nikitin2011ewt} suggest that the phase shift parameter is equal to $\delta_l \approx \pi / 4$ at $q_\parallel = 0$.
The resonance mode detected in graphene ribbons \cite{Ju2011gpf, Yan2013dpm} is evidently the $l = 1$ mode.
Probing $q_\parallel \neq 0$ modes in ribbons with conventional optics is challenging and has not been done is graphene
[It may be possible with a grating coupler \cite{Demel1991odp}.]
On the other hand, working with graphene disks,
one can effectively access the quantized values $q_\parallel = m / R$,
where $m$ is the azimuthal quantum number.
The observed mode \cite{Yan2012ist, Yan2012pcg, Fang2013gth} is evidently the dipolar one, $m = l = 1$,
which has the highest optical weight.
An additional mode that appears in both in ribbons and disks is the 
\textit{edge} plasmon.
We will talk about it at the end of this section where we discuss the effects of magnetic field.

The correspondence between the ribbon and bulk plasmon dispersions enables one to also verify the  $|N|^{1/4}$-scaling predicted by Eq.~\eqref{eqn:omega_p_q}.
This has been accomplished by electrostatic gating 
of graphene micro-ribbons immersed in ionic gel \cite{Ju2011gpf} and monitoring their resonance frequency.

Plasmons in graphene are believed to strongly interact with electrons.
Using the ARPES \textcite{Bostwick2007qdi, Bostwick2007rog, Bostwick2010oop}
observed characteristic departure of the quasiparticle dispersion from linearity near the Dirac point energy accompanied by an additional dispersion branch.
These features, discussed in more detail in Sec.~\ref{sec:e-ph-pl}, were interpreted in terms of plasmarons:
bound states of electrons and plasmons \cite{Lundqvist1967sps}.
\textcite{Walter2011esa, Walter2011eso} demonstrated that the details of the plasmaron spectrum are sensitive to dielectric environment of graphene.
\textcite{Carbotte2012eop} proposed that plasmaron features can be detected in near-field optical measurements,
which allow one to probe the IR response at momenta $q \gg \omega / c$.

Complementary insights on the interaction between plasmons and quasiparticles have been provided by the STS\@.
Based on the gate dependence of the tunneling spectra, \textcite{Brar2010ooc} distinguished phonon and plasmon effects on the quasiparticle self-energy.

Plasmons in graphene strongly interact with surface phonons of polar substrates such as SiC, SiO$_2$, and BN\@.
Dispersion of mixed plasmon-phonon modes in graphene on SiC was investigated experimentally using high-resolution EELS \cite{Liu2008pda, Liu2010pps, Koch2010spp} and modeled theoretically by \textcite{Hwang2010ppc}.
Theoretical dispersion curves  
\cite{Fei2011ino} for graphene on SiO$_2$ are shown in the inset of Fig.~\ref{fig:3.3.1}b.
The dispersion characteristic of mixed plasmon-phonon modes in nanoribbons measured via far-field IR spectroscopy was reported in \cite{Yan2012pcg, Yan2013dpm}.

In the near-field IR nanoscopy study of graphene micro-crystals on SiO$_2$ \cite{Fei2011ino}
the oscillator strength of the plasmon-phonon surface modes was shown to be significantly enhanced by the presence of graphene,
Fig.~\ref{fig:3.3.1}a.
The strength of this effect can be controlled by electrostatic doping, in agreement with
theoretical calculations \textcite{Fei2011ino}.

Imaging of plasmon propagation in real-space \cite{Fei2012gto, Chen2012oni} [Fig.~\ref{fig:3.3.1}(left)] have led to the first
direct determination of both real and imaginary parts of the plasmon momentum $q_p = q_p' + i q_p''$ as a function of doping.
In terms of potential applications of these modes,
an important characteristic is the confinement factor $\lambda_p / \lambda_0$,
where $\lambda_p = 2 \pi / q_p'$ is the plasmon wavelength and $\lambda_0 = 2 \pi c / \omega$
to the wavelength of light in vacuum.
Experimentally determined confinement factor in exfoliated graphene  \textcite{Fei2012gto} was $\sim 65$ in the mid-IR spectral range $\omega \approx 800\,\mathrm{cm}^{-1}$.
According to Eqs~\eqref{eqn:omega_p_q}, the scale for the confinement is set by the inverse fine-structure constant,
$\lambda_0 / \lambda_p = (\kappa / 2)(\hbar c / e^2)(\hbar \omega / \,|\mu|\,)$,
with stronger confinement achieved at higher frequencies.
The propagation length of the plasmons $\sim 0.5\, \lambda_p = 100$--$150\,\mathrm{nm}$ is consistent with
the residual conductivity $\sigma' \approx 0.5 e^2 / h$ measured by the
conventional IR spectroscopy \cite{Li2008dcd}.
Possible origins of this residual conductivity have already been discussed above, Sec.~\ref{sec:direct}.
In confined structures one additional mechanism of plasmon damping is scattering by the edges \cite{Yan2013dpm}.
Despite the observed losses, the plasmonic figures of merits demonstrated by \textcite{Fei2012gto, Chen2012oni} compare well against the benchmarks set by noble metals.
Even though surface plasmons in metals can be confined to scales of the order of tens of $\mathrm{nm}$,
their propagation length in this regime is plagued by giant losses and does not exceed $0.1 \lambda_p \sim 5\, \text{nm}$ for Ag/Si interface \cite{Jablan2009pig}.
This consideration has not been taken into account in a recent critique of graphene plasmonics \cite{Tassin2012aco}.
Further improvements in the figures of merits are anticipated for graphene with higher electronic mobility.
The key forte of graphene in the context of plasmonics is the control over the plasmon frequency and propagation direction \cite{Mishchenko2010gpi, Vakil2011tou} by gating.

The properties of graphene plasmons get modified in the presence of a transverse magnetic field $B$.
The magnetoplasmon dispersion is obtained from Eq.~\eqref{eqn:q_p} by replacing $\sigma$ with the longitudinal conductivity $\sigma_{x x}$.
For instance, instead of the Drude model~\eqref{eqn:sigma_Drude},
one would use its finite-$B$ analog,
the Drude-Lorentz model~\cite{Ashcroft1976ssp},
which yields another well-known dispersion relation \cite{Chiu1974pot}
\begin{equation}
\label{eqn:omega_mp}
\omega_{mp}(q) = \sqrt{\omega_p^2(q) + \omega_c^2}\,.
\end{equation}
This magnetoplasmon spectrum is gapped at the 
cyclotron frequency
$\omega_c = e B / m c$ defined through the effective mass $m$ [Eq.~\eqref{eqn:m}].
Equation~\eqref{eqn:omega_mp} is valid at small enough $B$ where Landau quantization can be ignored.
At large $B$, quantum treatment is necessary.
In the absence of interactions, 
the magnetoplasmon gap at $q = 0$ is given by $E_{n + 1} - E_n$,
the energy difference between the lowest unoccupied $n + 1$ and the highest occupied $n$ Landau levels.
Unlike the case of a 2DEG, where the Kohn's theorem holds,
renormalization of the Fermi velocity by interactions directly affects the cyclotron gap.
This many-body effect has been
observed by magneto-optical spectroscopy, Sec.~\ref{sec:Landau}.

Probing finite-$q$ magnetoplasmons optically is possible via the finite-size effects,
such as the mode quantization in graphene disks.
As known from previous experimental \cite{Glattli1985dhe, Mast1985obe,
Demel990ndr, Demel1991odp, Kukushkin2003ore},
numerical \cite{Eliasson1986mms}, and analytical \cite{Volkov1988eml}
studies of other 2D systems,
the single plasmon resonance at $B = 0$ splits into two.
The upper mode whose frequency increases with $B$ can be regarded the bulk magnetoplasmon with $q \approx 1 / R$,
where $R$ is the disk radius.
The lower mode whose frequency drops with $B$ can be interpreted as the edge magnetoplasmon,
which propagates around the disk in the anti-cyclotron direction.
Both the bulk-like and the edge-like modes have been detected by the IR spectroscopy of graphene disk arrays \cite{Yan2012ist, Yan2012pcg}.
Additionally, in epitaxial graphene with a random ribbon-like microstructure,
the $B$-field induced splitting of the Drude peak into a high- and a low-frequency branch was observed and interpreted in similar terms~\cite{Crassee2012itp}.
The distinguishing property of the edge magnetoplasmon is chirality:
its the propagation direction is linked to that of the magnetic field.
This property has been verified in graphene systems by time-domain spectroscopy \cite{Petkovic2013cdv, Kumada2013pti},
which also allowed extraction of the edge magnetoplasmon velocity.

Other interesting properties of magnetoplasmons,
such as splitting of the classical magnetoplasmon dispersion~\eqref{eqn:omega_mp} into multiple branches have been predicted theoretically \cite{Roldan2009cmd, Goerbig2011epg}
and their similarities and differences with the 2DEG case have been discussed.
These effects still await their experimental confirmation.


\subsection{Phonons} 
\label{sec:phonons}



\begin{figure}
\includegraphics[width=3.2in]{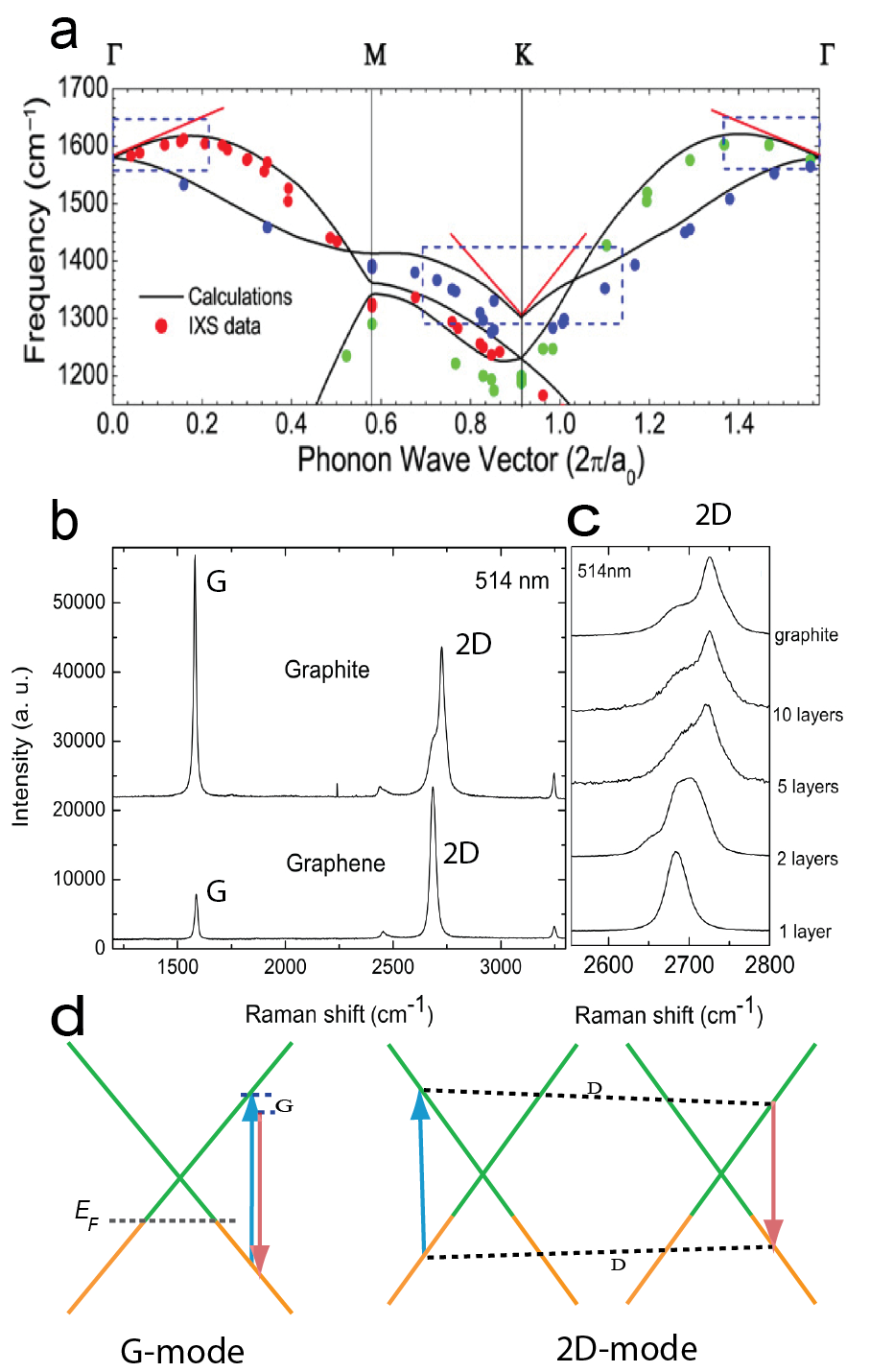}
\caption{\label{fig:3.4.1}
(Color online)
Phonon dispersion and Raman spectroscopy of graphene.
(a) Calculated phonon dispersion of SLG \cite{Piscanec2004kaa} (symbols) compared with the experimental data for graphite \cite{Mohr2007pdo} (lines).
(b) Raman spectra of graphene and graphite measured at $514\,\text{nm}$ laser excitation showing the $G$ and the $2D$ Raman peaks \cite{Ferrari2006rso}.
(c) The evolution of the $2D$ Raman peak with the number of graphene layers \cite{Ferrari2006rso}.
(d) Schematics of the $G$-mode and the $2D$-mode Raman scattering processes.} 
\end{figure}

Raman spectroscopy is the most widely used tool for probing optical phonons in graphene and related materials \cite{Ferrari2006rso, Ferrari2007rso, Dresselhaus2010poc, Dresselhaus2012rsc}.
Quantitative studies of the Raman modes can provide rich information on graphene electron-phonon interaction, electronic structure,
as well as on graphene layer thickness, edges, doping, and strain.
Because graphene has the same $s p^{2}$ bonding and hexagonal carbon lattice,
its phonon band-structure is almost identical to that in graphite.
Figure~\ref{fig:3.4.1}a shows calculated dispersion of the optical phonon branches in graphene (lines) \cite{Piscanec2004kaa} as well as the experimental data of graphite (symbols) \cite{Maultzsch2004pdi}.
One feature of these dispersions is the discontinuity in the frequency derivative at the $\Gamma$ and $\mathrm{K}$ points in the highest optical branches.
This discontinuity known as the Kohn anomaly arises from the unusual electron-phonon coupling in graphitic materials \cite{Piscanec2004kaa}.

Figure~\ref{fig:3.4.1}b displays typical Raman spectra of SLG and graphite.
They show the same qualitative Raman modes, with the two most prominent features being the $G$-mode ($\approx 1580\,\text{cm}^{-1}$) and the $2D$-mode ($\approx 2700\,\text{cm}^{-1}$, also known as $G'$ mode).
The other weak but very informative Raman feature is the $D$-mode ($\approx 1350\,\text{cm}^{-1}$).
The lineshape of $2D$ mode is very different in SLG, MLG, and graphite
(Fig.~\ref{fig:3.4.1}c) \cite{Ferrari2006rso}.
As illustrated in Fig.~\ref{fig:3.4.1}d,
the $G$-peak arises from Raman scattering of the $\Gamma$-point phonon.
The $2D$-peak, on the other hand, is a two-phonon process involving emission of two $\mathrm{K}$-point optical phonons.
The $D$-peak is a double resonance process like the $2D$-peak.
It requires structural defects to relax the momentum conservation constraint.

\begin{figure*}
\includegraphics[width=7.0in]{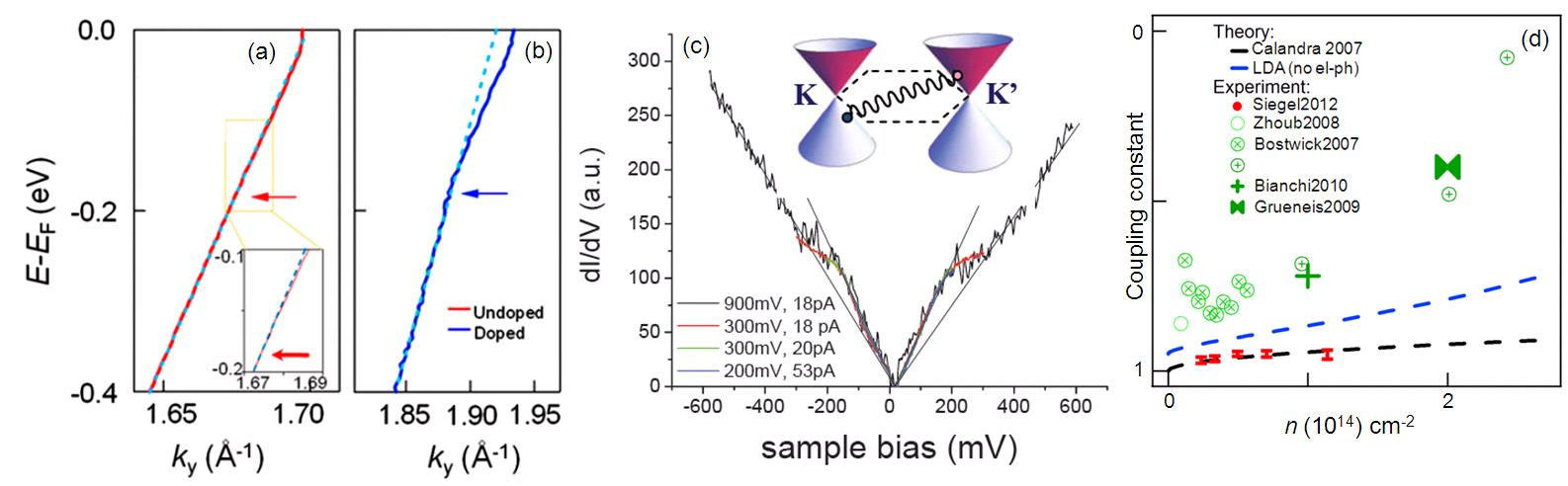}
\caption{\label{fig:el-ph}
(Color online)
The ARPES dispersion along the $\Gamma$-$\mathrm{K}$ direction for (a) lightly  doped (solid red line) and (b) heavily doped (solid blue line) graphene \cite{Siegel2011mbi}.
The dashed lines are guides to the eye.
The arrows indicate the kinks in the curves due to the el-ph interaction.
(c) Zero field STS tunneling spectra \cite{Li2009sts}.
The spectra are approximately linear except around the phonon energy.
(d) The ARPES electron-phonon coupling constant as a function of the carrier density for graphene grown on different substrates: SiC \cite{Zhou2008kaa, Bostwick2007qdi, McChesney2007meo, McChesney2010evh}, Ir~(111) \cite{Bianchi2010epc, Grueneis2009esa}, and Cu \cite{Siegel2012epc}.
The results of \textit{ab initio} calculations of \textcite{Calandra2007epc} without (blue dashed line) and with (black dashed line) el-ph interaction included are shown for comparison.}
\end{figure*}

A detailed theory of the $G$-mode Raman signal was presented in \textcite{Basko2008ioc, Basko2008tor, Basko2009eei, Basko2009cot}.
The capability of controlling the electron Fermi energy through electrical gating
helped to elucidate electron-phonon interactions \cite{Yan2007efe, Pisana2007bot, Malard2008ood, Das2008mdb} and the quantum interference between different intermediate excitation pathways \cite{Chen2011cil, Kalbac2010tio}.
The frequency and linewidth of the Raman $G$-mode reflect the energy and lifetime of the optical phonon at the $\Gamma$ point.
The $\Gamma$-point phonon experiences Landau damping by particle-hole excitations if its energy exceeds $2 |\mu|$ (see Fig.~\ref{fig:Excitations}).
As a result, the parameters of the Raman $G$-mode depend on the carrier concentration, as demonstrated experimentally \cite{Yan2007efe, Pisana2007bot}.
The $G$-mode Raman shows a reduced damping and a blue shift when the Fermi energy is larger than one half of the phonon energy,
so that the phonon decay pathway into electron-hole pairs gets blocked.
%
When the Fermi energy in graphene is increased further, some of the intermediate electronic transitions necessary for Raman scattering become blocked.
This reduces destructive interference among different pathways and increases the $G$-mode signal \cite{Chen2011cil}.

The Raman scattering that gives rise to the $2D$ mode involves emission of two BZ-boundary phonons close to the $\mathrm{K}$-point.
Being a two-phonon process, it still has large intensity, which is explained by the triple-resonance mechanism (Fig.~\ref{fig:3.4.1}d), where every intermediate step involves a resonant electronic excitation \cite{Basko2008tor, Basko2009eei}.
Due to smallness of the phonon energy compared with the incident photon energy $\hbar\omega$,
the momenta $\mathbf{k}$ of the intermediate electron states are restricted to $\hbar\omega \approx E(\mathbf{k})$,
where $E(\mathbf{k})$ is the electron dispersion (Sec.~\ref{sec:single}).
The phonon momentum (relative to a $\mathrm{K}$-point phonon) then equals $2 (\mathbf{k} - \mathbf{K})$.
Consequently, phonons and intermediate electronic transitions with specific momentum can be excited by varying incident photon energy for $2D$ Raman modes.
This allows one to map the dispersion of both the phonon and the electrons.

Once the phonon dispersion is known, Raman scattering can be used to probe electronic band-structure changes with a fixed laser excitation.
For example, it can distinguish SLG, BLG, and MLG due to their different electronic dispersions \cite{Ferrari2006rso}.
In BLG and MLG there are several conduction and valence bands (Sec.~\ref{sec:BLG}).
Hence, valence electrons at more than one momentum $\mathbf{k}$ can satisfy the $\hbar \omega = E(\mathbf{k})$ relation.
This leads to an apparent broadening and asymmetry of the $2D$ Raman peaks for BLG and MLG,
compared to those for SLG \cite{Ferrari2006rso}.

The Raman $D$-mode (short for the defect-mode) requires the existence of atomically sharp defects to provide the required momentum matching to scatter a zone boundary phonon close to $\mathrm{K}$-point.
The intensity of the $D$-peak is used to characterize the sample quality of graphene \cite{Malard2009rsi, Dresselhaus2010poc, Ferrari2007rso}.
The $D$-mode is also useful for probing graphene edges, which can be considered as line defects.
Experiments show that the $D$-peak is indeed the strongest at graphene edges \cite{Graf2007srr, Gupta2009pge, Casiraghi2009rso}, and that the $D$-mode intensity is at maximum for light polarization parallel to the edge and at minimum for the perpendicular polarization \cite{Casiraghi2009rso}.
For ideal edges, theory predicts that the $D$-mode Raman peak intensity is zero for zigzag edges but large for armchair ones \cite{Casiraghi2009rso}.
In addition to the effects discussed above, the intensity and frequency of Raman peaks also depend on the substrate \cite{Wang2008rso, Ni2008uso, *Ni2009uso, Lee2008rso, Berciaud2008pti}, temperature \cite{Calizo2007tdo}, and strain \cite{Yu2008rmi, Proctor2009hpr, Mohiuddin2009usi, Huang2009psa} through their effects on the phonon dispersion and electron Fermi energy.

\subsection{Electron-phonon and electron-plasmon interaction} 
\label{sec:e-ph-pl}

The energies and lifetimes of charge carriers in graphene are significantly affected by interactions with plasmons and phonons.
The electron-phonon (el-ph) interaction results in a variety of novel phenomena discussed in Sec.~\ref{sec:phonons}.
The ARPES has been used to probe the signature of the el-ph interaction in the electronic spectra
of graphene \cite{Bostwick2007qdi, McChesney2007meo, McChesney2010evh, Zhou2008dft}
via the measurement of the quasiparticle velocity $v$.
The el-ph coupling constant is usually defined by $\lambda = v_0 / v - 1$ \cite{Ashcroft1976ssp}.
However, electron-electron (el-el) interaction also contributes to
velocity renormalization (Secs.~\ref{sec:interaction} and \ref{sec:renormalization}).
Hence, thus defined $\lambda$ gives a good estimate of el-ph coupling only if el-el interaction is screened, which is the case for graphene on a metallic substrate \textcite{Siegel2011mbi}.

The el-ph interaction in graphene strongly depends on the carrier concentration,
as shown in Fig.~\ref{fig:el-ph}a,b.
\textcite{Siegel2011mbi} have reported a large reduction of $\lambda$ for quasi-free-standing graphene with $E_{F}$ close to the Dirac point $E_{D}$.
The overall reduction of the el-ph interaction can be reproduced by theoretical calculations \cite{Park2007vra}.
However, to account for the fine features of the quasiparticle dispersion,
the el-el interaction has to be included \cite{Siegel2011mbi, Zhou2008dft, Lazzeri2008iot}.
At high doping $\lambda$ appears to be enhanced,
reaching values $\lambda \sim 2$,
and strongly anisotropic, similar to what is observed in graphite \cite{Zhou2006lee, Leem2008eol, Park2008epi} and in the intercalated compound CaC$_{6}$ \cite{Valla1999mbe}.
\textcite{Calandra2007eso} argued these effects
result from distortion of the graphene bands that hybridize with a new Ca-related band.
On the other hand, \textcite{Park2008vhs} suggested that
the anisotropy of $\lambda$ comes from the nonlinear band dispersion of the graphene bands at high doping.

\begin{figure}
\includegraphics[width=3.1in]{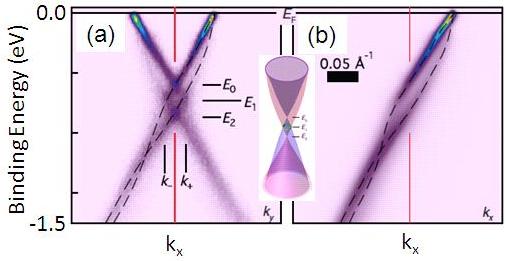}
\caption{\label{fig:3.6.2}
(Color online)
The ARPES dispersion of doped ($N = 8 \times 10^{10}\,\text{cm}^{-2}$) graphene perpendicular (a) and parallel (b) to the $\Gamma$-$\mathrm{K}$ direction \cite{Bostwick2010oop}.
The dashed black lines are guides to the eye for the dispersion of the hole and plasmaron bands; the solid red line goes through the Dirac point.
The inset shows a schematic of the renormalized spectrum in the presence of plasmarons.}
\end{figure}

From a high resolution ARPES study \textcite{Zhou2008kaa}
concluded that the electron-phonon coupling is dominated by the following phonon modes:
$A_{1g}$ phonon
at approximately $150\pm 15\,\text{meV}$ near the BZ corner,
$E_{2g}$ phonon ($\sim 200\, \text{meV}$) at the zone center,
and the out-of-plane phonon at $60\, \text{meV}$.
Among these, the $A_{1g}$ phonon is the one that mostly contribute to $\lambda$ and mainly responsible for the kinks in the ARPES and in the tunneling spectra \cite{Li2009sts},
see Fig.~\ref{fig:3.6.2}b,c.
The contribution of a specific phonon mode to $\lambda$ can also be determined
by studying how the Raman signal varies as a function of the applied magnetic field. 
These magneto-Raman studies focused on the $E_{2g}$ phonon~\cite{Faugeras2009tep, Faugeras2011mrs}, as the $A_{1g}$ phonon is Raman inactive in high quality graphene samples.

The origin of the large discrepancy (Fig.~\ref{fig:el-ph}d) between theoretically predicted and experimentally measured values of $\lambda$ is debated.\footnote{See \textcite{Bianchi2010epc, Bostwick2007rog, Park2008vhs, Park2007vra, McChesney2007meo, McChesney2010evh, McChesney2008sca, Zhou2008kaa, Calandra2007epc, Filleter2009fad, Grueneis2009esa}.}
\textcite{Siegel2012epc} found a good agreement with the theory (Fig.~\ref{fig:el-ph}d) using the bare velocity $v_0$ measured for graphene grown on Cu where
the el-el interaction is expected to be screened.

Electron-plasmon interaction is also believed to play an important role in renormalizing the band structure of graphene.
\textcite{Bostwick2007qdi, Bostwick2007rog, Bostwick2010oop} have argued that this interaction is responsible for the anomalous departure from the linear dispersion observed in epitaxial graphene grown on the Si face of SiC.
\textcite{Bostwick2010oop} have provided evidence (Fig.~\ref{fig:3.6.2}) for a well-resolved plasmaron band in the ARPES spectra of a ``freestanding'' graphene sample in which
hydrogen has been intercalated between graphene and SiC
to make negligible the interaction between the two.
The plasmaron of momentum $\mathbf{k}$ is a bound state of a hole with momentum $\mathbf{k} + \mathbf{q}$ and a plasmons of momentum $-\mathbf{q}$ \cite{Lundqvist1967sps}.
Theoretical calculations \cite{Polini2008pat} within the $GW$ approximation predict that
the plasmaron band appears at finite charge densities.
Its energy separation from the primary quasiparticle band is proportional to $\mu$
with a coefficient that depends on the Coulomb interaction strength $\alpha$,
which in turn depends on the dielectric environment of graphene.
Quantitative aspects of these calculations were disputed by \textcite{Lischner2013pos} who included vertex corrections neglected in the $GW$ scheme.
Compared to \textcite{Polini2008pat}, for the same $\alpha$
\textcite{Lischner2013pos} find a broader plasmaron peak at a smaller separation from the primary band,
which appears to be in a better agreement with the experiments of \textcite{Bostwick2010oop}.
 
No evidence of the plasmaron band has been reported in samples where decoupling of graphene from the buffer later was achieved by either gold or fluorine intercalation \cite{Walter2011hpd, Starodub2011ipo}.
This has been attributed to a stronger dielectric screening by the buffer layer.
An alternative interpretation of the apparent nonlinearity of the Dirac spectrum of graphene on SiC invokes a substrate-induced band gap \cite{Zhou2007sib, Zhou2008kaa, Zhou2008oot, Benfatto2008sso, Kim2008ooa}, see Sec.~\ref{sec:substrate} below.

%
%

\section{Induced effects} 
\label{sec:Induced}

\subsection{Inhomogeneities and disorder}
\label{sec:Inhomogeneities}


Intentional and unintentional doping by charged impurities plays a very important role in the electronic phenomena of graphene.
It is unclear if there is a single dominant source of unintentional doping even for most studied type of samples:
exfoliated graphene on SiO$_2$.
In addition to adsorbates from the ambient atmosphere,
doping could also result from
charged defects in SiO$_2$ \cite{Adam2007sct, Wehling2007mdo, Schedin2007doi, Zhou2008mti, Coletti2010cna}
lithographic residues \cite{Dan2009iro}, and metal contacts \cite{Connolly2010sgm}.

The dopants introduce not only a change in the average carrier concentration
but also charge inhomogeneities and scattering.
Near the graphene neutrality point inhomogeneities of either sign can arise, which are often
referred to as the electron-hole puddles \cite{Geim2007rg}.
Thus, even at the neutrality point the graphene is always locally doped.
This is a qualitative explanation for nonvanishing conductivity \cite{Geim2007rg, Tan2007mos, Chen2008cis} and TDOS \cite{Martin2008ooe}.
A more detailed model \cite{Adam2007sct, Hwang2007ctt, Shklovskii2007smo, Rossi2008gsg}
invokes a system of conducting electron-rich and hole-rich regions separated by $p$-$n$ junctions \cite{Cheianov2007tfo, Zhang2008nsa}.
The transport involves percolation through the $p$ and $n$ regions aided by tunneling across the junctions \cite{Cheianov2007rrn, DasSarma2011eti}.
Many elements of this semiclassical model
hark back to the earlier studies of two-dimensional \cite{Efros1993dos, Fogler2004nsa} and three-dimensional \cite{Efros1984epd} electron systems in semiconductors.
However, the puddle model may not be quantitatively reliable for graphene.
The correlation length of the density inhomogeneities is typically very short.
For SLG on SiO$_2$ it was consistently estimated to be
of the order of $20\,\mathrm{nm}$
using several complementary
scanned probes microscopy techniques \cite{Deshpande2011icd, Luican2011qll, Berezovsky2010ict, Martin2008ooe}.
A typical electron-hole ``puddle'' is also too small to contain even a single charge \cite{Martin2008ooe}.
Therefore, the inhomogeneities in question may be better described as
quantum interference patterns of disorder-scattered electron waves rather than large semiclassical puddles.
The situation may change if Coulomb interactions among electrons and impurities is screened. 
The crossover to the semiclassical regime is predicted to occur \cite{Fogler2009npo} 
for graphene on a substrate of high dielectric constant $\kappa \gg 1$.
%
%
Suppression of density inhomogeneities in one graphene layer due to screening by a nearby second layer
has been invoked to explain the observed localization transition in graphene-hBN-graphene structures \cite{Ponomarenko2011tmi}.

The inhomogeneities may also be induced by elastic strain and ripples \cite{Brey2008eic, Guinea2008msa, Gibertini2010edd}.
Electron density inside the highly strained graphene bubbles \cite{Bunch2008iam, Levy2010sip, Georgiou2011gbw} is undoubtedly inhomogeneous.
However, the relation between strain and electron density is nonlocal.
Indeed, no local correlations between the carrier density in graphene and 
the roughness of SiO$_2$ substrate is evident in scanned probe images \cite{Martin2008ooe, Deshpande2011icd, Zhang2009oos}.

The hypothesis that unintentional doping is caused by impurities trapped under graphene is supported by some micro-Raman experiments showing that proximity to the SiO$_{2}$ substrate results in increase of carrier density \cite{Berciaud2008pti, Ni2009pci, Bukowska2011rso}.
Yet other micro-Raman measurements \cite{Casiraghi2009rso} have not observed such correlations.

\begin{figure*}
\includegraphics[width=5.0in]{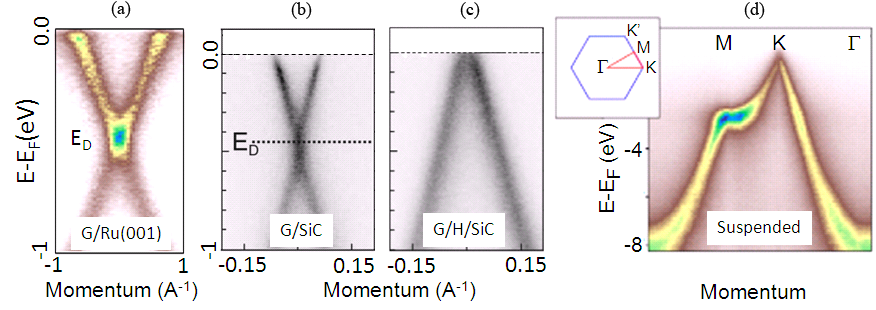}
\caption{\label{fig:4.1.1}
(Color online)
The ARPES intensity along (a)-(c) the direction orthogonal to $\Gamma$--$\mathrm{K}$ (d) the path $\Gamma$--$\mathrm{M}$--$\mathrm{K}$--$\Gamma$.
Panels (a), (b) and (c), and (d) are adopted from, respectively, \cite{Enderlein2010tfo},
\cite{Riedl2009qfs}, and \cite{Knox2011mar}.
}
\end{figure*}

Charge inhomogeneities can be reduced by either removing the substrate \cite{Du2008abt, Knox2011mar} or using a high-quality hBN substrate \cite{Dean2010bns}.
The random charge fluctuations of exfoliated graphene on hBN
are at least two orders of magnitude smaller than those on SiO$_2$
according to the STM studies \cite{Xue2011stm, Decker2011lep}.
(However, in such structures periodic charge oscillations may appear instead of random ones, see Sec.~\ref{sec:moire}.)
These random fluctuations are on par with the values estimated from transport data for free-standing graphene \cite{Du2008abt}.
The electronic mobility of graphene on hBN approaches $\sim 10^5\,\mathrm{cm}^2/\mathrm{Vs}$ implying the
mean-free path of several hundreds nm \cite{Du2008abt, Dean2010bns}.

Although detrimental for transport properties,
impurities can play a role of elementary perturbations that help reveal useful physical information.
We can give two examples.
First, disorder-induced LDOS fluctuations seen in STS \cite{Rutter2007sai, Zhang2009oos}
reveal the dominant momenta for inter- and intra-valley scattering and therefore
shed light on chirality and energy spectrum of the quasiparticles.
Second, by utilizing ionized Co adatoms one can study screening properties of graphene.
The screening cloud surrounding the adatoms was shown
to have a qualitatively different profile depending on the total charge of the adatom cluster.
In the sub-critical case this profile is governed essentially by the linear response dielectic constant of graphene.
Theoretical modeling of the STS spectra \cite{Brar2011gci, Wang2012mdq} suggests the enhanced value $\epsilon \approx 3.0$ of this constant,
which is indicative of many-body interactions \cite{Sodemann2012ict}. 
In the super-critical case \cite{Wang2013oac} sharp resonances in the local DOS appear,
which is the hallmark of a nonlinear screening with intriguing analogy to ``atomic collapse'' of super-heavy elements.

\subsection{Substrate-induced doping} 
\label{sec:substrate}

Metallic substrates induce a strong doping of graphene, which is readily seen by the ARPES (Fig.~\ref{fig:4.1.1}a).
The chemical potential $\mu = E_{F} - E_D$ measured with respect to the Dirac point ranges from approximately $0.5\,\text{eV}$ for Cu~(111) \cite{Gao2010ego} and Cu films \cite{Siegel2012epc, Walter2011eso} to $2\,\text{eV}$ for other transition metals,
such as Ni~(111) \cite{Nagashima1994eso, Dedkov2008rei, Varykhalov2008eam}, Ru~(0001) \cite{Himpsel1982abd, Enderlein2010tfo, Sutter2009}, and Co~(0001) \cite{Rader2009ita}.
An exception to this is graphene on Ir~(111) \cite{N'Diaye2006tdi, Pletikosic2009dca}, where the surface states of the substrate cause pinning of $\mu$ near zero.

Naively, graphene is $n$-doped if $W_G > W_M$ and $p$-doped otherwise,
where $W_G = 4.5\,\text{eV}$ is the work function of pristine graphene and $W_M$ is that of the metal.
In fact, the charge transfer is affected by chemical interaction between graphene and the metal and by their equilibrium separation \cite{Giovannetti2008dgw}.
The amount of charge transfer can be modified by intercalation.
Fluorine intercalation yields a large $p$-type doping of graphene \cite{Walter2011hpd}.
Hydrogen intercalation leads to decoupling of graphene from the substrate \cite{Riedl2009qfs},
as evidenced by the ARPES dispersions, 
Fig.~\ref{fig:4.1.1}(b) and (c), typical of suspended graphene, cf.~Fig.~\ref{fig:4.1.1}(d).
Similar effects can be obtained by Au intercalation \cite{Gierz2008ahd}.
When gold atoms are intercalated between graphene and a Ni~(111) substrate \cite{Varykhalov2008eam},
$\mu$ drops down to $25\,\text{meV}$, corresponding to the two orders of magnitude decrease in the
carrier concentration.

\subsection{Moir\'e patterns and energy gaps} 
\label{sec:moire}

When the lattice constants of the graphene layer and the substrate differ by a small relative amount $\delta$ and/or misoriented by an angle $\phi$ a moir\'e supelattice arises \cite{Marchini2007stm, N'Diaye2006tdi, Wang2008coo, Wintterlin2009gom}.
The electron dispersion in the presence of the moir\'e superlattice gets modified as a result of hybridization of the original Dirac cones with their replicas folded into a superlattice Brillouin zone (sBZ).
Such replicas have been seen in the ARPES spectra of graphene on Ir~(111) \cite{Pletikosic2009dca}
although they may also be due to the final-state diffraction \cite{Sutter2009}.

\begin{figure*}
\includegraphics[width=5.5in]{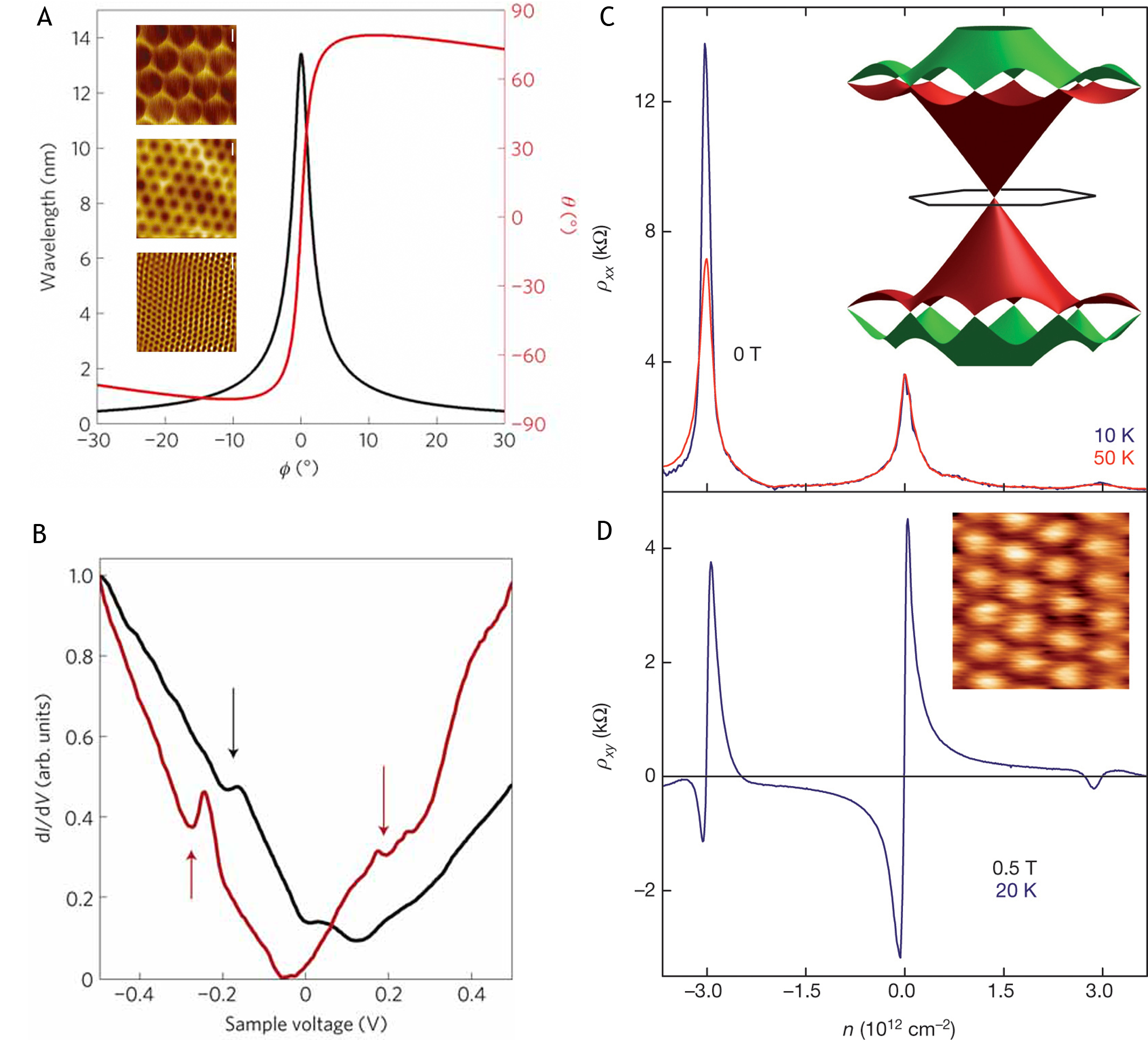}
\caption{\label{fig:Moire}
(Color online)
A. Moir\'e superlattice wavelength (black) and rotation (red) as a function of the angle between the SLG and hBN lattices.
Inset: STM topography images showing (top to bottom) $2.4$, $6.0$,
and $11.5\,\mathrm{nm}$ moir\'e patterns.
The scale bars in all the images are $5\mathrm{nm}$ tall.
B. The tunneling $dI/dV$ curves for samples with moir\'e wavelengths $9.0\,\mathrm{nm}$ (black) and $13.4\,\mathrm{nm}$ (red).
The dips in the $dI/dV$ curves marked by arrows occur at secondary Dirac points.
C. Longitudinal resistivity $\rho_{xx}$ of SLG on hBN as a function of carrier density.
Inset: one possible reconstruction of SLG spectrum.
D. The Hall resistivity $\rho_{xy}$ changes sign at high electron and hole doping, revealing well-isolated secondary Dirac points.
Inset: conductive atomic force microscope image of the moir\'e pattern.
The separation between the white spots is $11\,\mathrm{nm}$.
Panels A-B and C-D are adapted from \cite{Yankowitz2012esd} and \cite{Ponomarenko2013cdf},
respectively.
}
\end{figure*}

The most striking experimental manifestations of the moir\'e superlattice effects have recently been observed in SGL on hBN.
This system has $\delta = 1.8\%$, so that the moir\'e period can be as long as $14\,\mathrm{nm}$,
which can be easily imaged by scanned probes (Fig.~\ref{fig:Moire}, insets).
Dependence of the moir\'e period on the misorientation angle $\phi$ is very sharp,
Fig.~\ref{fig:Moire}A, so achieving large period requires precise alignment.

It has been predicted theoretically that at
the intersections of the replica and the main bands new Dirac points appear, Fig.~\ref{fig:Moire}C.
For the practical case of a weak superlattice potential,
these points have energy
\begin{equation}
E_D^m \simeq E_D \pm \frac{2\pi}{\sqrt{3}}\, \frac{\hbar v}{\Lambda}\,,
\label{eqn:moire_DP}
\end{equation}
where $\Lambda$ is the moir\'e period.
[For the opposite limit of strong modulation, see \cite{Brey2009ezm}.]
The extra Dirac points
are characterized by a modified and generally, anisotropic quasiparticle velocity \cite{Park2008ngo, Guinea2010bsa, Wallbank2013gms}.
The original Dirac point at the center of the sBZ remains gapless,
at least,
within the scope of generic theoretical models of the moir\'e superlattice which preserve sublattice symmetry.

First attempts to identify extra Dirac points in SLG/hBN structures were unsuccessful because these points were outside the experimental energy window \cite{Xue2011stm}.
In more recent experiments,
which utilized precisely aligned ($\phi < 0.5^\circ$) structures,
the new Dirac points are clearly evidenced by additional minima of the DOS measured by STS 
\cite{Yankowitz2012esd}, Fig.~\ref{fig:Moire}B.
The unmistakable signatures of the second-generation Dirac points in transport include peaks in longitudinal resistance and the sign change of the Hall resistance \cite{Ponomarenko2013cdf, Dean2013hba, Yankowitz2012esd, Yang2013egs}, Fig.~\ref{fig:Moire}C and D.
Additional Landau level (LL) fans emerging from these extra Dirac points are seen in magnetotransport \cite{Ponomarenko2013cdf, Dean2013hba, Hunt2013mdf} and the gate capacitance measurements.
The detailed structure of such LLs is predicted to be fractal,
as spectacularly illustrated by the iconic image of the ``Hofstadter butterfly'' \cite{Hofstadter1976elw}.
Experiments \cite{Hunt2013mdf}
in strong fields $B > 20\,\mathrm{T}$ demonstrate additional quantum Hall plateaus
and a large gap (tens of meV) at the neutrality point,
the physical origin of which remains to be understood.

Opening a band gap at the Dirac point is indeed the most often cited
effect that can enable wider applications of graphene.
Inducing gap by confinement in various graphene superstructures such as quantum dots, ribbons, \textit{etc.}, has proved to be problematic due to disorder effects.
Inducing a gap through graphene/substrate interaction seems an attractive alternative.
The most straightforward mechanism of the gap generation is breaking the sublattice symmetry of graphene \cite{Brey2006eso, Giovannetti2007sib, Nakada1996esi, Nilsson2007tta, Trauzettel2007sqi} 
For a hypothetical commensurate SLG/hBN structure, a band gap $\sim 50\,\mathrm{meV}$ was predicted \cite{Giovannetti2007sib, Slawinska2010egt}.
Theoretically, the gap can also be induced by hybridization of the two valleys \cite{Manes2007eat}. 
A small band gap in graphene can also be induced by a spin-orbit coupling of Rashba type \cite{Kane2005qsh}.
The curvature of the graphene sheet is predicted to enhance the Rashba splitting \cite{Huertas-Hernando2006soc, Kuemmeth2008cos}.

Existence of substrate-induced gaps have been indicated by many ARPES experiments.
A very wide spread of gap values has been reported, which remains unexplained.
One of the earliest ARPES studies \cite{Oshima1997ute} claimed the largest gap so far, $1.3\,\text{eV}$, for a ``soft'' SLG on TaC~(111).
Large gaps have also been reported for graphene on certain metallic substrates.
\textcite{Brugger2009coe} found $\sim 1\,\mathrm{eV}$ gap for SLG on Ru~(0001).
\textcite{Nagashima1994eso} observed $0.7$--$1.3\,\text{eV}$ gaps in SLG on Ni~(111) intercalated with alkaline metals.
For Ru~(0001) covered a with monolayer of gold \cite{Enderlein2010tfo} and for Ir~(111) substrates \cite{Starodub2011ipo} the gap is $0.2\,\text{eV}$.
The effect of intercalants is counterintuitive because these are metals to which graphene interacts weakly.
The gap for SLG on Cu is $0.3$--$0.4\,\text{eV}$.
\textcite{Zhou2007sib} made a case for the $0.26\,\text{eV}$ gap in epitaxial graphene on SiC.
As discussed in Sec.~\ref{sec:e-ph-pl},
a competing interpretation of these data is in terms of plasmarons
\cite{Bostwick2007qdi, Bostwick2007rog, Bostwick2010oop}.
Comparable gaps were found for other semiconducting substrates  \cite{Siegel2011mbi, Walter2011eso}.
For graphene on graphite the gap of $20\,\text{meV}$ \cite{Li2009sts, Siegel2011mbi} was reported.
The STS \textcite{Kawasaki2002dal} observed a $0.5\,\text{eV}$  for SLG/single-layer hBN/Ni~(111) structure.

The largest Rashba splitting $13\pm 3\,\text{meV}$ has been reported for graphene on magnetic substrate Ni~(111) intercalated by Au  \cite{Varykhalov2008eam}.
The mechanism behind this enhancement is still unknown.
No Rashba splitting has been observed on another magnetic substrate, Co~(0001) intercalated by Au \cite{Rader2009ita}.
Although intrinsic spin-orbit (SO) coupling is also responsible for the opening a band gap, $\Delta_{\mathrm{SO}}$, in pure SLG it is predicted to be extremely small, e.g., $10^{-3}$--$10^{-2}\,\mathrm{meV}$
\cite{Kane2005qsh}.
A broken symmetry at the interface of two SLG can somewhat amplify this gap~\cite{Schmidt2010ese}.
Impurities in graphene resulting in $s p^3$ type deformation of the flat graphene are also predicted to enhance $\Delta_{SO}$ up to $7\,\mathrm{meV}$~\cite{CastroNeto2009iis}.
In fact, a recent experimental study on hydrogenated graphene revealed a drastically enhanced $\Delta_{\mathrm{SO}}$ of $2.5\,\mathrm{meV}$~\cite{Balakrishnan2013ces}.
Alternatively, the interactions of charge carriers in graphene with heavy atoms such as In and Tl adsorbed on graphene are predicted to enhance $\Delta_{\mathrm{SO}}$ up to $7\,\mathrm{meV}$ and $21\,\mathrm{meV}$, respectively~\cite{Weeks2011erq}.

\subsection{Elastic strain}
\label{sec:strain}

\begin{figure}
\includegraphics[width=2.4in]{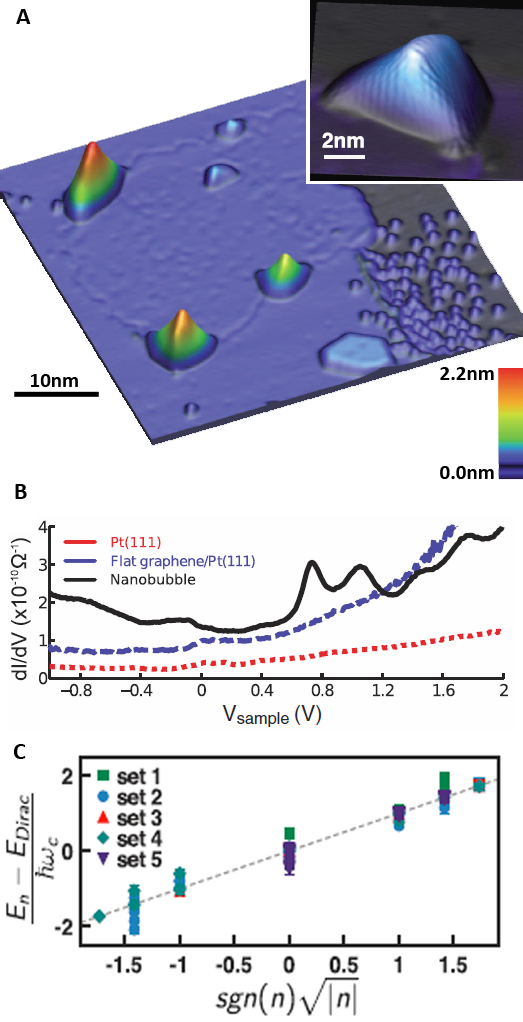}
\caption{\label{fig:4.2.1}
(Color online)
STM images and STS spectra taken at $7.5\,\text{K}$.
(A) Graphene monolayer patch on Pt(111) with four nanobubbles at the graphene-Pt border and one in the patch interior.
Residual ethylene molecules and a small hexagonal graphene patch can be seen in the lower right (3D $z$-scale enhanced $4.6\times$).
(Inset) High resolution image of a graphene nanobubble showing distorted honeycomb lattice resulting from strain in the bubble ($\text{max}\, z = 1.6\,\text{nm}$, 3D $z$-scale enhanced $2\times$).
(B) STS spectra of bare Pt(111), flat graphene on Pt(111) (shifted upward by $3 \times 10^{-11}\, \Omega^{-1}$), and the center of a graphene bubble (shifted upward by $9 \times 10^{-11}\, \Omega^{-1}$). The peaks in the graphene bubble spectrum indicate the formation of pseudo-Landau levels.
(C) Normalized peak energy versus $\text{sgn}\,(n) \sqrt{|n|}\,$ for peaks observed on five different nanobubbles follow expected scaling behavior (see text).
Adapted from~\textcite{Levy2010sip}.}
\end{figure}

A controlled uniaxial strain can be readily introduced into graphene by stretching the flexible substrate.
The strain modifies graphene phonon energy spectrum, which is effectively probed by Raman spectroscopy.
Under uniaxial strain the $G$ and $2D$ phonon bands display significant red shift proportional to the applied strain: a result of the anharmonicity of the interatomic potentials in graphene \cite{Ni2008uso, *Ni2009uso, Huang2009psa, Mohiuddin2009usi, Tsoukleri2009sag}.
Meanwhile the $sp^{2}$ bonds of graphene lengthen/shorten in the direction parallel/perpendicular to the strain axis.
This reduces the $C_{6}$ symmetry of the honeycomb lattice to $C_{2}$, and splits the doubly degenerate $G$ band into two singlet bands, $G^{+}$ and $G^{-}$, with normal modes perpendicular and parallel to the strain axis, respectively.
The polarization of Raman scattered light for the $G^{+}$ and $G^{-}$ modes is thus expected to depend on the direction of the strain axis relative to the crystal orientation: a conjecture verified by \textcite{Huang2009psa} and \textcite{Mohiuddin2009usi}.

Strain also introduces profound modifications to graphene electronic structure.
The defining topology feature of graphene electronic band, namely the degeneracy of conical electron and hole bands at Dirac points, is protected by the inversion symmetry of graphene lattice \cite{Hasegawa2006zmo, Kishigi2008ego, Wunsch2008dpe}.
A small perturbation in form of mechanical strain does not lift the degeneracy, but deforms the energy bands and shifts the Dirac points in both the energy and the momentum space.
The former is equivalent to a scalar potential also known as the deformation potential.
A general nonuniform strain generates a spatially varying dilation of the graphene lattice and therefore local ion density.
The deformation potential arises because the corresponding Coulomb potential is only partially screened by electrons \cite{Suzuura2002pae, Kim2008gaa, Guinea2008gfi}.
Next, shifting the Dirac point in the $k$-space
away from the $\mathrm{K}$ ($\mathrm{K}'$) points \cite{Pereira2009seo, Farjam2009cob, Ni2008uso, *Ni2009uso, Dietl2008nmf}.
is analogous to the effect induced by an external magnetic field applied perpendicular to the graphene plane.
One can parametrize the mechanical strain by a gauge field $\mathbf{A}$ \cite{Iordanskii1985dal, Kane1997ssa, Sasaki2005leg, Morpurgo2006isl, Katsnelson2007gnb, Fogler2008pfa} and define the pseudomagnetic field $B_s = \bm{\nabla} \times \mathbf{A}$.
The strain-induced $\mathbf{A}$ and $B_s$ have opposite signs at two valleys $\mathrm{K}$ and $\mathrm{K}'$,
so that the time-reversal symmetry is preserved.

It is possible to engineer a special nonuniform strain for which $B_s$ is approximately constant in a finite-size region.
If such pseudomagnetic field is strong enough, it can lead to Landau quantization and quantum Hall-like states \cite{Guinea2010ega, Guinea2010gqp}.
%
In a recent experiment \cite{Levy2010sip}, such an unusual Landau quantization has been observed in highly strained graphene nanobubbles.
The strain arises upon cooling because of a mismatch in the thermal expansion coefficients of graphene and the Pt substrate (Fig.~\ref{fig:4.2.1}A).
The pseudo-Landau levels, manifested as local density of states peaks, are probed by STS (Figs.~\ref{fig:4.2.1}B and C).
Their energies follow the theoretically predicted $\text{sgn}(n) \sqrt{|n|}$ behavior with a gigantic $B_s \sim 300\,\text{T}$, see Fig.~\ref{fig:4.2.1}C.

\subsection{Photo-induced effects} 
\label{sec:photo-induced}

Optical spectroscopy with ultrafast laser excitation pulses provides a unique tool to probe the dynamic evolution of electrons and phonons in graphene, including the cooling of the non-equilibrium quasiparticle plasma through electron-electron and electron-phonon interactions and the relaxation of hot optical phonons.
These processes are not only of fundamental interest due to the unusual electronic structure in graphene, but also important for technological applications of high-field electronics and nonlinear photonic devices \cite{Xia2009ugp, Bao2009alg, Zhang2010gml, Sun2010gml, Bonaccorso2010gpa}.

\begin{figure}
\includegraphics[width=3.3in]{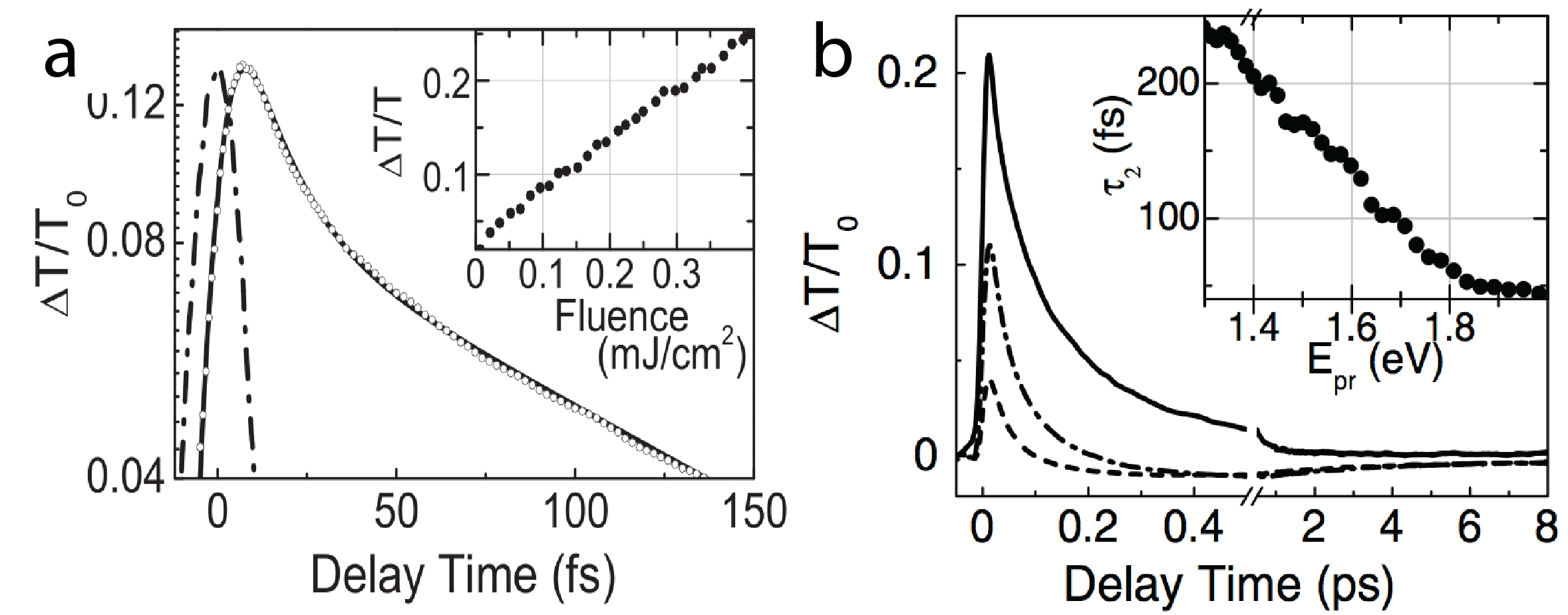}
\caption{\label{fig:4.4.1}
Ultrafast dynamics of excited electrons in graphene.
(a) Spectrally integrated transmission change as a function of pump-probe delay (open circles).
Solid line is the numerical fit, and dash-dotted line is cross correlation of pump and probe pulses.
The inset shows linear dependence of the maximum transmission change on the absorbed pump fluence.
The transmission increase is due to photo-induced Pauli blocking of interband transitions.
The ultrafast decay is due to thermalization between the electrons/holes and optical phonons.
(b) Transient transmission changes at probe photon energies of $1.24\,\text{eV}$ (solid), $1.55\,\text{eV}$ (dash-dotted) and $1.77\,\text{eV}$ (dashed) for short and long delays.
The slower decay at picosecond time scale is due to equilibration with acoustic phonons \cite{Breusing2009ucd}.}
\end{figure}

Response of graphene to a pulsed laser excitation has been studied by several complementary ultrafast spectroscopy techniques.
For example, pump-probe IR/visible spectroscopy~\cite{Sun2008uro, Dawlaty2008mou, Newson2009uck, Huang2010uta} and pump-probe THz spectroscopy \cite{George2008uop} has been employed to track the time evolution of optical absorption and transmission by graphene.
Ultrafast photoluminescence~\cite{Lui2010upf, Liu2010nbp, Stoehr2010fol} has been used to monitor light emission by non-equilibrium electron gas.
Time-resolved Raman spectroscopy~\cite{Yan2009trr, Chatzakis2011tdo}
has explored generation and decay of hot optical phonons.
\textcite{Breusing2009ucd} studied graphite using $7\,\mathrm{fs}$ $1.55\,\mathrm{eV}$ pump pulses and a broadband probe pulses with energy spectrum from $1.2$ to $2\,\mathrm{eV}$.
Figure~\ref{fig:4.4.1}a shows the observed increase of transmission during the first $150\,\mathrm{eV}$ with sub-$10\,\mathrm{fs}$ time resolution.
This phenomenon is attributed to partial Pauli blocking of the optical transition by photo-excited electron-hole pairs.
The change in transmission scales linearly with the pump influence and decays with two time constants of $13\,\mathrm{fs}$ and $100\,\mathrm{fs}$.
The former characterizes electron-electron interactions,
which cause energy redistribution within the conduction and valence band,
as well as relaxation of occupation factors by Auger processes.
The second time constant describes interaction of quasiparticles with optical phonons.
The emission of optical phonons with energy $\approx 0.2\,\mathrm{eV}$ cools down electron-hole plasma.
Once its temperature drops below this number,
emission of optical phonons becomes ineffective.
Eventual equilibration of the electron and lattice temperatures is achieved by
emission of acoustic phonons on a time scale of $\sim 2\,\mathrm{ps}$.
 
The decay dynamics of graphene in the $8\,\text{ps}$ temporal range is shown in Fig.~\ref{fig:4.4.1}b with three different probe photon energies.
In addition to the fast decay processes described above, it shows a slower relaxation process with a time constant of $1.4\, \text{ps}$.
This picosecond time scale reflects partial thermalization between the hot electron/holes and optical phonons with the acoustic phonons in graphene.
We note that the photo-induced transmission change can become negative at certain probe photon energies at longer delay.
This is because optical excitation not only leads to Pauli blocking of interband transitions, but also increased high frequency absorption from intraband transitions.
At longer pump-probe delay, the intraband absorption can dominate over the Pauli blocking effects at some probe photon energies.
Similar photo-induced transmission decreases have also been observed in optical pump-THz probe measurements, where photo-induced intraband transitions always dominate \cite{George2008uop, Strait2011vsc, Wright2009sno}.

Ultrafast photoluminescence monitors the light emission from the highly non-equilibrium electrons after femtosecond pump excitations \cite{Lui2010upf, Liu2010nbp, Stoehr2010fol}.
Broad light emission across the visible spectral range ($1.7$--$3.5\, \text{eV}$) was observed with femtosecond near-IR laser excitation, where the incident photon has an energy of $1.5\,\text{eV}$.
This unusual blue-shifted photoluminescence exhibits a nonlinear dependence on the laser fluence, and it has a dominant relaxation time within $100\,\text{fs}$.
This nonlinear blue-shifted luminescence was attributed to recombination of hot electron-hole plasma generated right after the femtosecond excitation.

In addition to the electron dynamics, researchers were able to probe the phonon dynamics specifically using time-resolved Raman spectroscopy \cite{Chatzakis2011tdo, Yan2009trr}.
Such studies show a decay lifetime of $2.5\,\text{ps}$ for the BZ-center $G$-mode phonons.
This time scale corresponds to the cooling of the optical phonons through anharmonic coupling to acoustic phonons, and the $2.5\,\text{ps}$ time constant is similar to that obtained in pump-probe transmission spectroscopy.

\section{Bilayer and multilayer graphene} 
\label{sec:BLG}

There has been a rapidly increasing interest in graphene systems with more than one layer (for an early review, see \textcite{Nilsson2008epo}).
The electronic structure of BLG and MLG is distinctly different from that of SLG.
These differences give rise to many new phenomena,
ranging from a tunable bandgap \cite{McCann2006lld, McCann2006agi, Castro2007bbg, Oostinga2008gii, Zhang2009doo} to strongly correlated ground states
\cite{Feldman2009bss, Bao2010moa, Weitz2010bss, Mayorov2011ids, Velasco2012tso}.
Unfortunately, space limitations and the open debate on the nature of these low-energy states
\cite{Nilsson2006eei, Min2008pmi, Barlas2010aec, Nandkishore2010dsa, Nandkishore2010qah, Nandkishore2011pke,
Vafek2010mbi} do not permit us to describe them in any detail.
In this short section we confine ourselves to discussing ``higher'' energy properties of these materials
that have been measured by ARPES and by optical spectroscopy.

We begin with discussing the quasiparticle dispersion of BLG.
For the Bernal stacked BLG, which is the most energetically favorable structure,
it is conventionally described by means of five parameters.
They include four hopping integrals $\gamma_0 = 3.0\,\text{eV}$, $\gamma_1 = 0.41\,\text{eV}$, $\gamma_3 = 0.3\,\text{eV}$, $\gamma_{4} = 0.15\,\text{eV}$, and also
the on-site energy shift $\Delta' = 0.018\,\text{eV}$ (Fig.~\ref{fig:1.2.1}B).%
\footnote{For a discussion of these numerical values and comparison with graphite,
see \textcite{Li2009bsa, Zhang2008dot, Kuzmenko2009iso, Kuzmenko2009gti}.}
When BLG is subject to an electric field due to external gates or charged impurities, the sixth parameter,
a scalar potential $\pm \Delta / 2$ on the two layers must be included.
The interlayer bias $\Delta = e E d$ is given by the product of the layer separation $d = 0.335\,\mathrm{nm}$ and the 
electric field $E$ between the layers.

\begin{figure*}
\centering
\includegraphics[width=7.20in]{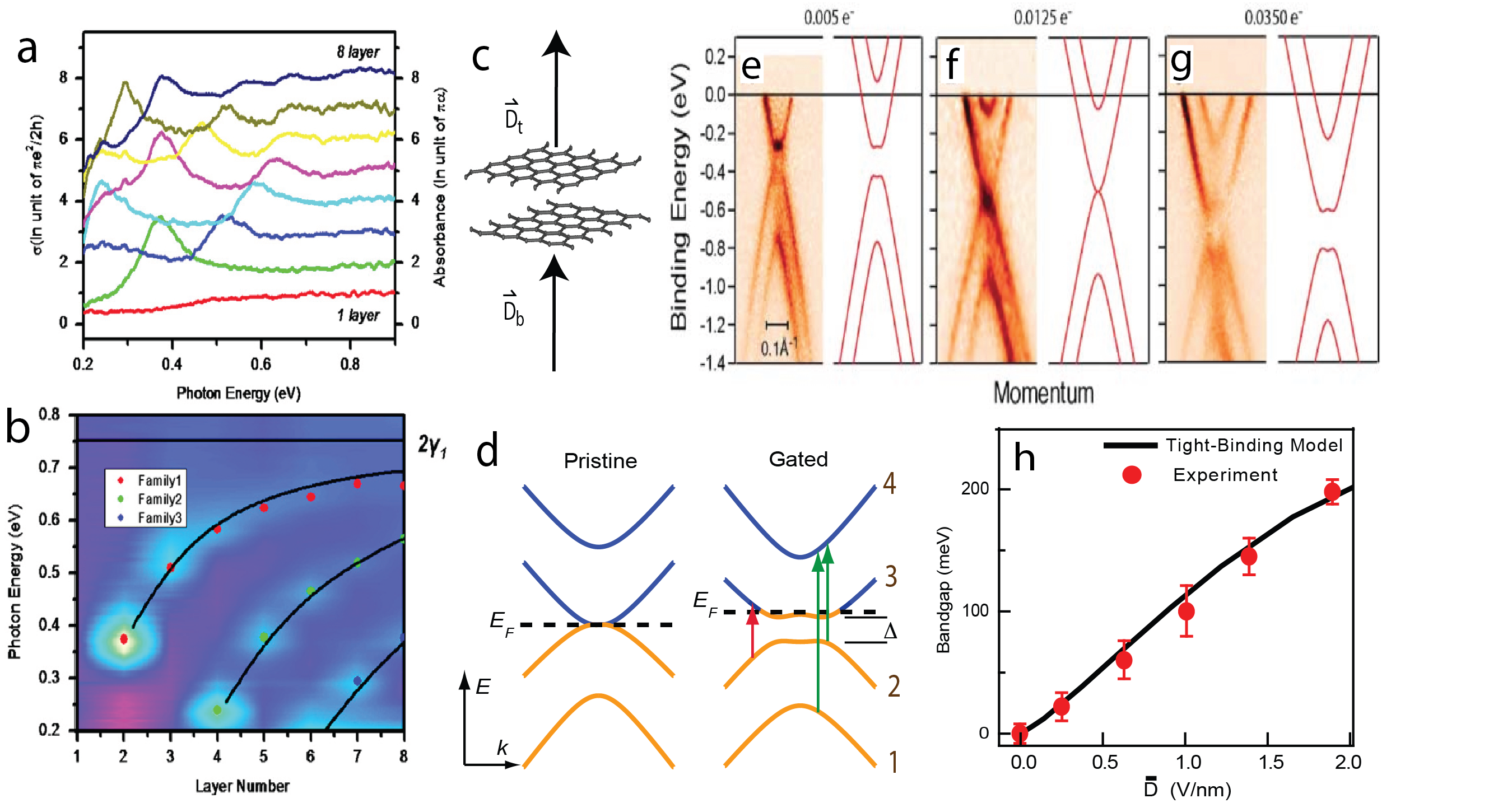}
\caption{\label{fig:5.1.1}
(Color online)
(a) IR conductivity spectra of MLG with layer number $L = 1, 2, \dots,  8$.
(b) A contour plot of the IR conductivity per layer as a function of photon energy and $L$.
The dots identify the position of the peaks in experimental IR conductivity.
These transition energies follow well-defined energy-scaling relations predicted by the zone-folding model (solid curves) \cite{Mak2010teo}.
(c) An illustration of the Bernal stacked BLG with electrical field above and below it.
(d) Electronic structure of a pristine and a gated BLG.
Arrows indicate allowed optical transitions.
(e)-(g) ARPES data showing the evolution of an induced bandgap in electronic structure of epitaxial BLG on SiC with chemical adsorbed Ca atoms \cite{Ohta2006ces}.
(h) Electrical field dependence of the induced bandgap in BLG measured through IR absorption spectroscopy (symbols)
\cite{Zhang2009doo}.}
\end{figure*}

The BLG has four atoms in the unit cell, and so the electron spectrum consists of four bands.
The two outer (lowest and highest energy) bands are hyperbolic,
with the extremal values at approximately $\pm \gamma_1$ reached at the BZ corners.
The shape of the two inner bands is more intricate.
At high energies they are nested with the outer bands.
At low energies their dispersion depends on the relation between the gap $\Delta$ and the hopping integral $\gamma_3$,
which causes the trigonal warping.
At $\Delta \gg \gamma_3 / (\gamma_0 \gamma_1)$, the trigonal warping is a small effect.
The bands are shaped as sombreros, e.g.,
the conduction band has a local maximum at energy $\Delta / 2$ at $q = 0$ and a local minimum  --- the bottom of the sombrero --- at the ring $|\mathbf{q}| \approx \Delta / \sqrt{2}\, \hbar v_0$, see Fig.~\ref{fig:5.1.1}d (right) \cite{McCann2006lld}.
As $\Delta$ is decreased, the trigonal warping of the bottom of the sombrero becomes more and more pronounced.
In the absence of the interlayer bias, $\Delta = 0$, the parabolic extrema split into the four conical Dirac points.
This reconstruction is an example of the Lifshitz transition.
The linear rather than parabolic shape of the bands in the symmetric BLG is supported by
the linear-$T$ dependence of low-temperature electric conductivity in extremely clean suspended BLG
\cite{Mayorov2011ids}.
However, in less perfect samples, the quadruple Dirac cones structure is smeared by disorder. It is
therefore common to approximate the inner bands by hyperboloids touching at a point,
see Fig.~\ref{fig:5.1.1}d (left).

In order to vary $\Delta$, one has to apply an external electric field to BLG.
This can be achieved experimentally through electrostatic gating or doping.
If $D_{t}$ and $D_{b}$ are electric displacement fields on the two sides of BLG (Fig.~\ref{fig:5.1.1}c),
then the interlayer electric field $E$ is determined by their mean $\bar{D} = (D_{b} + D_{t}) / 2$.
Notably, $E$ is smaller than $\bar{D}$ due to screening effects.
Calculations within the self-consistent Hartree approximation \cite{McCann2006lld}
predict a factor of two or so reduction in typical experimental conditions.

The difference $D = D_{b} - D_{t}$ of the displacement fields
produces a net carrier doping, and so the Fermi energy shift (Fig.~\ref{fig:5.1.1}d).
Unless $D_t$ and $D_b$ are precisely equal or precisely opposite,
the modification of the band gap $\Delta$ and the shift of the Fermi energy $E_F$ occur simultaneously
\cite{McCann2006lld, McCann2006agi, Castro2007bbg}.

The control of electronic structure of BLG was first revealed in ARPES studies of potassium-doped epitaxial graphene on SiC \cite{Ohta2006ces}.
Figures~\ref{fig:5.1.1}e-g display the evolution of the ARPES spectra with doping.
As prepared, BLG is $n$-type doped.
This corresponds to a finite $D_b$ and zero $D_{t}$, leading to a nonzero bandgap (Fig.~\ref{fig:5.1.1}e).
Potassium adsorption generates a finite $D_{t}$.
When its value is the same as $D_{b}$, one obtains an electron doped gapless BLG (Fig.~\ref{fig:5.1.1}f).
With further increase in potassium doping, the bandgap reappears (Fig.~\ref{fig:5.1.1}g).
Tuning of BLG electron structure can be also achieved via coupling to
different substrates \cite{Siegel2010qmg}.

Complimentary insights on the band structure of BLG has been provided through IR spectroscopy.
There is a total of six inter-band optical transitions possible in this material.
The near-perfect nesting of two conductions bands results in a strong absorption peak at mid-IR energy $\gamma_1$,
when the transition between them is activated by $n$-type doping.
A refined estimate of $\gamma_1 = 0.40\pm 0.01\,\mathrm{eV}$ has been obtained by monitoring the
lineshape and position of this peak in a gated BLG structure as a function of the gate voltage and modeling
these spectra theoretically \cite{Zhang2009doo}.
Similarly, the $p$-type doping activates transition between the two valence bands.
From slight differences of $p$- and $n$-type spectra,
the electron-hole symmetry breaking parameters $\gamma_4$ and $\Delta'$ have been inferred.
The parameters obtained from the IR experiments are corroborated by those
derived from the Raman spectroscopy \cite{Malard2007pte} and the capacitance measurements of the
TDOS \cite{Henriksen2010mot}.

The bandgap tuning by electrical gating was demonstrated using IR spectroscopy
through monitoring the gate-induced change in other
three transitions shown in Fig.~\ref{fig:5.1.1}d by arrows \cite{Mak2009ooa, Kuzmenko2009dot, Zhang2009doo}.
The dependence on the bandgap on the mean displacement field $\bar{D}$ (Fig.~\ref{fig:5.1.1}h)
was found to be in agreement with the theory \cite{McCann2006lld}.

On the other hand, observing the predicted bandgap value from electrical transport measurements has been challenging.\footnote{See \textcite{Szafranek2010eoo, Xia2010gfe, Castro2007bbg, Oostinga2008gii}.}
Gated BLG typically exhibits an insulating behavior only at $T < 1\,\text{K}$, suggesting a very narrow gap \cite{Oostinga2008gii}.
This is because electrical transport is extremely sensitive to defects and impurities,
and a very high quality graphene is required to reach the intrinsic BLG behavior.
Recent transport studies \cite{Xia2010gfe} however demonstrate transport gaps closer to those obtained through IR spectroscopy.

Electron-phonon coupling in gated BLG shows up in tunable electron-phonon Fano resonances \cite{Tang2010atp, Kuzmenko2009gti, Cappelluti2010psa}.
There is a host of other effects that originate from unique gate-tunable electronic structure in BLG that
have been predicted theoretically and are amenable to spectroscopic studies,
e.g., a rich Landau level spectrum structure \cite{Zhang2011mcb}.
However, we must leave this topic now to at least briefly discuss MLG.

The electronic structure of MLG has been investigated experimentally using optical spectroscopy.
Figure~\ref{fig:5.1.1}a displays a set of IR absorption spectra from $L = 1$ to $8$ layer graphene over the photon-energy range of $0.2$--$0.9\,\text{eV}$ \cite{Mak2010teo}.
At energies in the range $\gamma_1 \ll \hbar\omega \ll \gamma_0$
the MLG is expected to behave in the first approximation as a stack of uncoupled SLG, each
possessing the universal optical conductivity $\sigma_0$, Eq.~\eqref{eqn:sigma_0}.
Indeed, at energies above $0.8\,\text{eV}$, the measured optical absorption scales linearly with $L$.
However, at lower energies, the absorption becomes highly structured and distinct for different $L$.
The evolution of the absorption spectra as a function of $L$ can be visualized from the false color plot
Fig.~\ref{fig:5.1.1}b in which
the principle transition energies are marked by the solid curves.
The shape of these curves can be understood through zone-folding of the graphite band-structure.
In particular, a gapless band is present if $L$ is odd and absent if it is even \cite{Mak2010teo}.

\section{Outlook}
\label{sec:outlook}

A wealth of spectroscopic data analyzed in this review has provided a panoramic picture of electronic phenomena in graphene.
The concept of 2D Dirac quasiparticles offers a unifying description of the gross features revealed in all spectroscopic and transport probes.
At the same time, pronounced and reproducible deviations from the predictions of noninteracting models of graphene have been documented.
An outstanding challenge for future research is probing these many-body effects using specimens of record-high electron mobility where the role of disorder is further reduced.

Due to space limitations, we have not been able to cover some topics at all, e.g.,
nanostructured graphene, spin phenomena, graphene at ultra-high doping, or fractional quantum Hall effect in graphene~\cite{Goerbig2011epg}.
We also covered some others, e.g., BLG and MLG in insufficient detail.
These topics are being actively explored, and the consensus is still being reached (for review, see \cite{McCann2013epb}).
In addition, a particularly interesting class of materials for future research are hybrid multilayer structures and superlattices assembled from graphene and other ultrathin atomic crystals, such as hBN, MoS$_2$, \textit{etc.}\cite{Geim2013vdw}.

Besides fundamental research, graphene and its spectroscopy have inspired a number of applications.
For example, the spectroscopic studies motivated the development of novel experimental tools and methods compatible with the architecture of gatable devices.
Novel scanning spectroscopies have advanced by exploiting the unique aspect of graphene that it is unobstructed by other interfaces.
Controlled modification of graphene properties has been demonstrated through using elastic strain,
interactions with the substrate, adatoms, and/or other graphene layers.
These new experimental approaches are expected to find applications in other areas of condensed matter physics.
Examples of viable device concepts spawned by
photon-based spectroscopies include compact (passive) optical components, photodetectors and bolometers, and
saturable absorbers.
In addition, standard plasmonics figures of merit show competitiveness or even superiority of graphene as a plasmonic medium compared to more seasoned metal-based technologies.
An unresolved question is whether graphene is suitable for achieving population inversion and lasing.

\section*{Acknowledgments}

D.~B., M.~F., and F.~W. acknowledge support from ONR
under Grant No.~N0014-13-0464.
The work at UCSD is also supported by DOE-BES under Contract No.~DE-FG02-00ER45799, by AFOSR Grant No. FA9550-09-1-0566, by
NSF Grant No.~DMR-1337356, by ARO Grant No. W911NF-13-1-0210,
and also by UCOP and FENA.
Additionally, F.~W. is supported by DOE-BES under Contracts
No.~DE-SC0003949 and No.~DE-AC02-05CH11231.
A.~L. acknowledges support from the Novel $sp^2$-bonded Materials Program
at Lawrence Berkeley National Laboratory, funded by the
DOE Office of Basic Energy Sciences, Materials Sciences and
Engineering Division under Contract No. DE-AC02-05CH11231.
Y.~Z. is supported by NSF of China through
Grant No.~11034001 and and MOST of China through Grant
No.~2011CB921802.

%

\end{document}